\pdfoutput=1
\documentclass[journal=jacsat,manuscript=article]{achemso}

\usepackage[numbers]{natbib}
\usepackage[version=3]{mhchem} 
\usepackage{graphicx}
\usepackage{svg}
\usepackage[export]{adjustbox}
\usepackage{wrapfig}
\usepackage[font=small,labelfont=bf]{caption}
\usepackage{gensymb}
\usepackage{multicol}
\usepackage{setspace}
\usepackage[normalem]{ulem}
\author{Jonathan W. P. Zajac}
\affiliation{Department of Chemistry, University of Minnesota, Minneapolis, MN 55455, USA}

\author{Praveen Muralikrishnan}
\affiliation{Department of Chemical Engineering and Materials Science, University of Minnesota, Minneapolis, MN 55455, USA}

\author{Idris Tohidian}
\affiliation[MTU]
{Department of Chemical Engineering, Michigan Technological University, Houghton, MI 49931, USA}

\author{Xianci Zeng}
\affiliation[UMA]
{Department of Chemical Engineering, University of Massachusetts Amherst, MA 01003, USA}

\author{Caryn L. Heldt}
\affiliation[MTU]
{Department of Chemical Engineering, Michigan Technological University, Houghton, MI 49931, USA}

\author{Sarah L. Perry}
\affiliation[UMA]
{Department of Chemical Engineering, University of Massachusetts Amherst, MA 01003, USA}

\author{Sapna Sarupria}
\email{sarupria@umn.edu}
\affiliation{Department of Chemistry, University of Minnesota, Minneapolis, MN 55455, USA}
\affiliation{Chemical Theory Center, University of Minnesota, Minneapolis, MN 55455, USA}

\date{\today}

\title{\noindent\LARGE{\textbf{Flipping Out: Role of Arginine in Hydrophobic Interactions and Biological Formulation Design}}}

\begin{document}
\doublespacing
\clearpage

\begin{abstract}
Arginine has been a mainstay in biological formulation development for decades. To date, the way arginine modulates protein stability has been widely studied and debated. Here, we employed a hydrophobic polymer to decouple hydrophobic effects from other interactions relevant to protein folding. While existing hypotheses for the effects of arginine can generally be categorized as either direct or indirect, our results indicate that direct and indirect mechanisms of arginine co-exist and oppose each other. At low concentrations, arginine was observed to stabilize hydrophobic polymer collapse via a sidechain-dominated direct mechanism, while at high concentrations, arginine stabilized polymer collapse via a backbone-dominated indirect mechanism. When adding partial charges to sites on the polymer, arginine destabilized polymer collapse. Further, we found arginine-induced destabilization of a model virus similar to direct-mechanism destabilization of the charged polymer, and concentration-dependent stabilization of a model protein similar to the indirect mechanism of hydrophobic polymer stabilization. These findings highlight the modular nature of the widely used additive arginine, with relevance in the design of stable biological formulations.
\end{abstract}

\section{Introduction}
Maintaining native protein structures in biological formulations poses a challenge, and is commonly addressed by strategic additive incorporation.\cite{jeong_analytical_2012, wang_protein_2005, patro_protein_2002}  Arginine stands out as a frequently employed additive in such formulations, spanning both therapeutic proteins\cite{arakawa_biotechnology_2007} and vaccines.\cite{hamborsky_epidemiology_2015, mistilis_long-term_2017} Arginine has been widely used as an aggregation suppressor, an agent for protein refolding, a cryoprotectant during lyophilization, and in protein purification.\cite{tsumoto_role_2004, tsumoto_review_2005, startzel_arginine_2018} Once hailed as a universal stabilizer, emerging studies paint a foggier picture of the effects of arginine. In some settings, the presence of arginine has accelerated the aggregation,\cite{smirnova_l-arginine_2013, shah_effects_2011, eronina_dual_2014} denaturation,\cite{xie_guanidine_2004, anumalla_counteracting_2019, arakawa_effects_2018} and inactivation\cite{meingast_arginineenveloped_2020, meingast_physiochemical_2021} of certain proteins and viruses.  Additional studies have found that arginine mechanisms are dependent on concentration.\cite{falconer_stabilization_2011, 
thakkar_excipients_2012, vagenende_protein-associated_2013, platts_controlling_2015} Hence, the existing literature on arginine reveals a lack of a cohesive understanding of its multi-faceted effects on protein stability.

In general, additive effects on protein stability are thought to be either direct or indirect. Direct mechanisms involve direct protein-additive interactions, while indirect mechanisms influence protein stability by modulating the surrounding solvent structure. It remains debated whether arginine acts primarily via a direct or indirect mechanism. Several studies called attention to direct interactions between arginine and aromatic residues,\cite{tsumoto_role_2004, arakawa_suppression_2007, shukla_interaction_2010} acidic residues,\cite{eronina_dual_2014, vagenende_protein-associated_2013, ng_mechanism_2024} and hydrophobic moieties.\cite{das_inhibition_2007, li_solubilization_2010} Other studies have proposed clusters of free arginine molecules in solution enable the crowding out of protein-protein interactions,\cite{schneider_investigation_2009, shukla_interaction_2010, schneider_arginine_2011, 
vagenende_protein-associated_2013} or alteration of hydration shell water dynamics.\cite{santra_analyzing_2021, santra_influence_2022} The wide range of observations related to the role of arginine on protein stability suggests that arginine harbors diverse, context-dependent mechanisms.

To elucidate the mechanisms through which arginine influences protein stability, this study focuses on its effects on hydrophobic interactions. Hydrophobic interactions are key in several biologically relevant phenomena, including protein folding and stability.\cite{kauzmann_factors_1959, tanford_contribution_1962, tanford_hydrophobic_1978, dill_dominant_1990, savage_lost_1993, hummer_new_2000, pratt_hydrophobic_2002, ben-naim_hydrophobic_2005,  chandler_interfaces_2005} Additives in solutions are known to modulate the strength of hydrophobic interactions, and in turn, the stability of proteins.\cite{timasheff_control_1993,ghosh_salt-induced_2005, athawale_osmolyte_2005, athawale_enthalpyentropy_2008, zangi_ureas_2009, canchi_cosolvent_2013, van_der_vegt_hydrophobic_2017} For example, simulation studies have shown trimethylamine N-oxide (TMAO) has a negligible effect on or strengthens hydrophobic interactions,\cite{athawale_osmolyte_2005, paul_hydrophobic_2008, macdonald_effects_2013, ganguly_hydrophobic_2016, su_effects_2018, folberth_small--large_2022} while these interactions are weakened in urea solutions.\cite{wallqvist_hydrophobic_1998,ikeguchi_molecular_2001,ghosh_salt-induced_2005, van_der_vegt_enthalpyentropy_2006, lee_does_2006, athawale_enthalpyentropy_2008, zangi_ureas_2009, shpiruk_effect_2013}  Indeed, the effects of TMAO and urea on hydrophobic interactions are consistent with their experimentally-observed roles as a protein stabilizer and denaturant, respectively.\cite{wang_naturally_1997, zou_molecular_2002, bennion_molecular_2003, canchi_cosolvent_2013}

Molecular dynamics (MD) simulations have provided valuable insights towards understanding these mechanisms as they relate to hydrophobicity. Several studies have highlighted the utilization of a hydrophobic polymer model for describing the role of solvent and additives on protein-like collapse. \cite{ten_wolde_drying-induced_2002, zangi_ureas_2009, nayar_cosolvent_2018, mondal_when_2013, athawale_enthalpyentropy_2008, jamadagni_how_2009} The use of a hydrophobic polymer model enables the decoupling of additive effects on hydrophobic vs other interactions, which is challenging in experiments. Additionally, comparison of arginine mechanisms on purely hydrophobic interactions allows indirect insights into experimentally-observed effects. For example, studies in which effects mirror what we observe from the hydrophobic polymer are likely to reflect a dominant mediation of hydrophobic interactions. In the present study, we utilize MD simulations of a hydrophobic polymer to characterize the effects of arginine on many-body hydrophobic interactions pertinent to protein stability, and contextualize these findings within larger-scale models for biological formulations.

Overall, we found arginine stabilizes hydrophobic polymer collapse at all concentrations under study. Interestingly, we discovered arginine sits on the edge of a mechanistic flip, balancing between direct- and indirect-dominated effects. As a consequence of this balance, we found subtle modulation of the polymer chemistry (via partial charge incorporation) changes arginine from a stabilizing additive to a destabilizer of polymer collapse. Consequently, in practical examples of formulation design, we observed arginine has variable effects on a model virus and protein.


\section{Methods}
\subsection{Hydrophobic Polymer System Setup and Molecular Dynamics Simulations}
We simulated a hydrophobic polymer in arginine solutions at different concentrations (Table S2). All simulations were performed using GROMACS 2021.4\cite{van_der_spoel_gromacs_2005, abraham_gromacs_2015} with the PLUMED 2.8.0 \cite{bonomi_plumed_2009,tribello_plumed_2014} patch applied. The hydrophobic polymer was modeled as a linear coarse-grained chain with 26 monomers, where each monomer represents a $\mathrm{CH_2}$ unit with Lennard-Jones parameters $\sigma= 0.373$ $\mathrm{nm}$ and $\epsilon= 0.5856$ $\mathrm{
 kJ/mol}$.\cite{athawale_osmolyte_2005} Box dimensions were defined such that 1.5 nm of space separated the fully elongated polymer from the nearest box edge. For simulations with no polymer, the same box dimensions were used. Arginine was modeled in accordance with a pH of 7, resulting in a protonated sidechain. An equal number of arginine molecules and Cl$^{-}$ atoms were added to the box until the desired concentration was reached. The TIP4P/2005\cite{abascal_general_2005} model was used to describe water, and the CHARM22 force field was used for arginine and Cl$^{-}$.\cite{brooks_charmm_2009}

All simulations were initially subject to energy minimization using the steepest descent algorithm. NVT equilibration was carried out for 1 ns at 300 K, followed by a 1 ns NPT equilibration at 300 K and 1 atm. During equilibration, temperature was controlled according to the V-rescale thermostat,\cite{bussi_canonical_2007} while pressure was controlled via the Berendsen barostat.\cite{berendsen_molecular_1984}  Following equilibration, NPT production runs were completed using the Nos\'e-Hoover thermostat\cite{evans_nosehoover_1985} and Parrinello-Rahman barostat.\cite{parrinello_polymorphic_1981} Production runs were completed for 20 ns for arginine/water systems, and between 50-250 ns per window for arginine/polymer/water replica exchange umbrella sampling (REUS) runs (Table 2). In all simulations, the Particle Mesh Ewald (PME) algorithm was used for electrostatic interactions with a cut-off of 1 $\mathrm{nm}$. A reciprocal grid of 42 x 42 x 42 cells was used with $\mathrm{4^{th}}$ order B-spline interpolation. A single cut-off of 1 $\mathrm{nm}$ was used for van der Waals interactions. The neighbor search was performed every 10 steps. Lorentz-Berthelot mixing rules\cite{lorentz_1881, berthelot1898melange} were used to calculate non-bonded interactions between different atom types, except for polymer-water oxygen interactions (see SI for details). 
 
\subsection{Replica Exchange Umbrella Sampling}
REUS\cite{sugita_multidimensional_2000} simulations were completed to sample the hydrophobic polymer conformational landscape in arginine solutions. The radius of gyration ($R_{g}$) of the hydrophobic polymer was used as the reaction coordinate to describe polymer folding/unfolding. 12 umbrella potential windows were centered between $R_{g}$ = 0.3 and $R_{g}$ = 0.9 nm, with a spacing of 0.05 nm. A force constant of $K$ = 5000 kJ/mol/nm$^2$ was used in all windows, with the exception of the window centered at $R_g$ = 0.45, which used $K$ = 1000 kJ/mol/nm$^2$ (see SI for details).

The potential of mean force (PMF) along the radius of gyration of the polymer was calculated as ${W(R_g) = -k_{B}T ln(P(R_g))}$. Biased probability distributions were reweighted according to the Weighted Histogram Analysis Method (WHAM).\cite{zhu_convergence_2012} The free energy of polymer unfolding ($\mathrm{\Delta G_u}$) was calculated according to:

\begin{equation}\label{unfolding_free_energy}
     \exp \left(\frac{\Delta G_{\text{u}}}{k_BT}\right)=\frac{\int_{R_{g,cut }}^{R_{g,\max}} \exp\left(\frac{-W\left(R_g\right)}{k_B T}\right) dR_g}{\int_{R_{g,\min}}^{R_{g,cut }} \exp\left(\frac{-W\left(R_g\right)}{k_B T}\right) dR_g}
\end{equation}

\noindent where $\mathrm{R_{g,cut}}$ was determined as the point between the folded and unfolded states where $
\frac{\partial W(R_{g})}{\partial R_{g}} = 0$. 

We decomposed the PMF into individual components to further investigate the role of arginine in polymer collapse. Following the methods outlined by several others,\cite{athawale_effects_2007,godawat_unfolding_2010,van_der_vegt_hydrophobic_2017,dasetty_advancing_2021} the PMF was decomposed as:
\begin{equation}
           W(R_g) = W_{vac} (R_g) + W_{cav} (R_g) + E_{pw}(R_g) + E_{pa} (R_g) + E_{pc} (R_g)
\label{eqn:decomp}
\end{equation}
$W_{vac} (R_g)$ captures intrapolymer degrees of freedom and was obtained from independent REUS simulations of the polymer in vacuum. $E_{pw} (R_g)$, $E_{pa} (R_g)$, and $E_{pc} (R_g)$ are average polymer-water, polymer-arginine, and polymer-chloride interaction energies, respectively. The remaining term is $W_{cav} (R_g)$, which is the cavitation component and quantifies the energetic cost of forming a cavity -- of the same size and shape as the polymer -- in the solution.

\subsection{Preferential Interaction Coefficients}
Distribution of arginine with respect to the polymer can be described via the preferential interaction coefficient ($\Gamma_{PA}$),\cite{scatchard_physical_1946, casassa_thermodynamic_1964, schellman_selective_1987}
\begin{equation}
    \Gamma_{PA}=-\left(\frac{\partial \mu_P}{\partial \mu_A}\right)_{m_P, T, P}=\left(\frac{\partial m_A}{\partial m_P}\right)_{\mu_A, T, P}
\end{equation}
where $\mu$ is the chemical potential, $m$ is the concentration and $W$, $P$, and $A$ refer to water, polymer, and an additive, respectively. This parameter is calculated in simulations using the two-domain formula\cite{inoue_preferential_1972, record_interpretation_1995, shukla_molecular_2009} given by:
\begin{equation}
    \Gamma_{PA}=\left\langle N_A^{\text {local }}-\left(\frac{N_A^{\text {bulk }}}{N_W^{\text {bulk }}}\right) N_W^{\text {local }}\right\rangle
\end{equation}
\noindent
where $N$ represents the number of molecules of a given species and angular brackets denote an ensemble average. The local and bulk domain was separated by a cutoff distance $R_{cut}$ from the polymer. $\Gamma_{PA}$ gives a measure of the relative accumulation or depletion of an additive in the local domain of the hydrophobic polymer, with $\Gamma_{PA} > 0$ indicating relative accumulation (preferential interaction) and $\Gamma_{PA} < 0$ indicating relative depletion (preferential exclusion).

\subsection{Hydrogen Bond Analysis}
Hydrogen bonds were calculated according to geometric criteria of a donor-acceptor distance of $r \leq$ 0.35 $\mathrm{nm}$ and donor-hydrogen-acceptor angle within 150\degree~\textless ~$\theta$ \textless ~180\degree.\cite{ferrario_molecular-dynamics_1990} Hydrogen bond existence correlation functions for water-water and arginine-water interactions were estimated according to:\cite{luzar_structure_1993,luzar_hydrogen-bond_1996,luzar_resolving_2000}
\begin{equation}
     C(\tau)=\left\langle\frac{\sum_{i,j}{h_{i j}\left(t_0\right)}{h_{i j}\left(t_0+\tau\right)}}{\sum_{i,j}{h_{i j}\left(t_0\right)^2}}\right\rangle
\end{equation}
\noindent
where $h_{i j}\left(t_0\right)$ is equal to 1 if there is a hydrogen bond between groups $i$ and $j$ at time $t_0$, and 0 if no hydrogen bond is present. An average over all possible values of the time origin $t_{0}$ was taken over the last 5 ns production simulations.

\subsection{Contact Coefficients}
Contact coefficients ($CC$) give a measure of excipient preference for interacting with specific residues. Here, we computed $CC$s as described by Stumpe and Grubm{\"u}ller\cite{stumpe_interaction_2007}:

\begin{equation}
     CC_X = \frac{N_{X-A}}{N_{X-W}}\frac{M_W}{M_A}
\end{equation}
\noindent where $N_{X-A}$ and $N_{X-W}$ are the number of atomic contacts of protein residue $X$ with an additive or water molecules, respectively. Atoms were defined to be in contact if any pair of heavy atoms were within a 0.35 nm cut-off. $CC$s are normalized by the total number of additive atoms ($M_A$) and water atoms ($M_W$) in solution. Values of $CC_X~ \textgreater ~1$ indicate preferential interaction with an additive for residue $X$, while $CC_X ~\textless ~1$ denotes preferential interaction with water.

\subsection{Protein and Virus Simulation Setup}
Porcine parvovirus (PPV) and hen egg white lysozyme (HEWL) were used as large-scale models for formulation design. To model PPV, we constructed a surface model using 15 monomers from the published crystal structure (PDB: 1K3V).\cite{simpson_structure_2002} This was achieved by selecting three reference $C_\alpha$ atoms from the 5-fold pore of the viral surface, and aligning vector normal to these points with the z-axis of the simulation box. Box dimensions were 24 x 24 x 12 nm, with the capsid surface model extending approximately 9 nm in the z-direction. Arginine molecules were added to the exterior of the capsid surface to reach a target concentration of 0.25 M, and NaCl atoms were distributed throughout the box to both neutralize the system and reach a target concentration of 0.15 mM. To prevent diffusion of arginine molecules into the capsid interior, a wall was placed at z = 0 nm, and the capsid surface was positioned on top of this wall (Fig. S12). A position restraint with a force constant of 1000 kJ/mol/nm$^{2}$ was applied to all capsid atoms within 1.5 nm from the wall. Capsid atoms between 1.5 nm and 3.0 nm from the wall were restrained with a force constant between 1-1000 kJ/mol/nm$^2$, with atoms further from the wall scaled to a lesser extent. Capsid atoms greater than 3.0 nm from the wall were left fully flexible. PPV simulations were carried out in the NVT ensemble -- first for 1 ns to reach the target temperature, followed 50 ns to allow the solvent to relax. 100 ns production runs were used for further analysis.

HEWL was modeled from an available crystal structure (PDB: 1LYZ),\cite{diamond_real-space_1974} and simulated in solution with 0.1, 0.2, 1.0, and 2.0 M arginine. HEWL simulations were carried out in the NVT ensemble for 1 ns to equilibrate to the target temperature, followed by 1 ns NPT simulations to reach the target density. HEWL production runs were completed for 100 ns in the NPT ensemble. For both HEWL and PPV, the force field, thermostats, and barostats matched those used for the hydrophobic polymer systems.

\subsection{Experimental Temperature Stability Studies}
Liquid samples of PPV in arginine solutions were prepared in triplicates and were put either in a heat block at 60 \degree C \cite{mi_thermostabilization_2020} or in a fridge at 4 \degree C as the control samples. 72 hours later, the titer of PPV was determined by 3-(4,5-dimethylthiazol-2-yl)-2,5-diphenyltetrazolium bromide (MTT) colorimetric cell viability assay.\cite{joshi_design_2024}  Differential scanning fluorimetry (DSF) was performed to compute the hydrophobic exposure temperature (HET) of HEWL in arginine solutions.\cite{seabrook_high-throughput_2013, shi_dsf_2013} More details for both experimental methods are included in the supporting information. 

\section{Results and Discussion}
\subsection{Arginine favors hydrophobic polymer collapse}

\begin{figure*}[!ht]
   \includegraphics[width=1\textwidth]{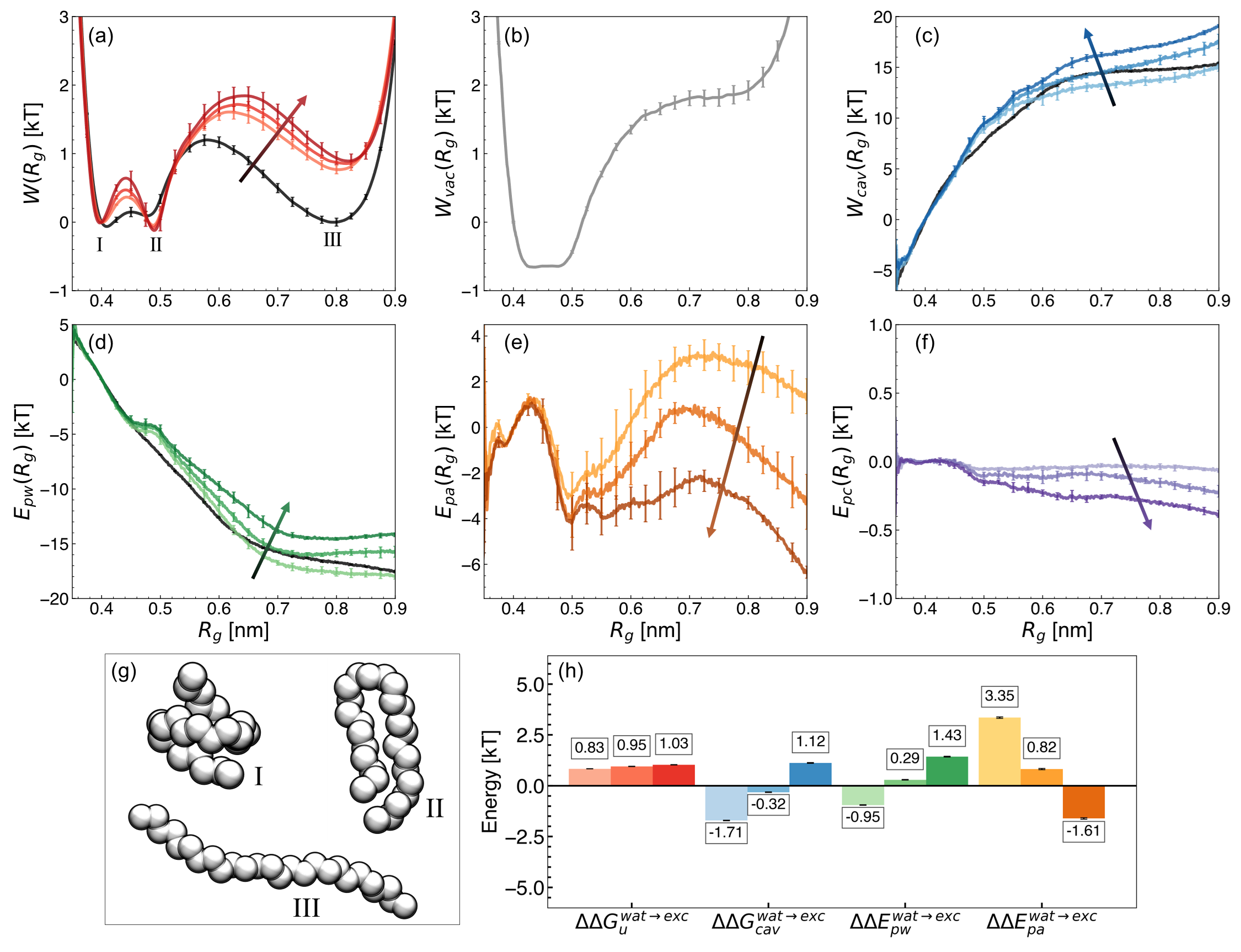}
    \caption{PMF decomposition in 0.0 M, 0.25 M, 0.5 M, and 1.0 M arginine solutions. (a) The PMF obtained along the $R_{g}$ reaction coordinate, $W(R_{g})$, (b) vacuum component, $W_{vac}$, (c) cavitation component, $W_{cav}$, (d) polymer-water interactions, $E_{pw}$, (e) polymer-arginine interactions, $E_{pa}$, and (f) polymer-chloride interactions, $E_{pc}$. (g) Representative configurations along the reaction coordinate as denoted in (a) as I, II, and III. (h) Changes in overall free energy of unfolding ($\Delta \Delta G_{u}$), cavitation contribution ($\Delta \Delta G_{cav}$), polymer-water interactions ($\Delta \Delta E_{pw}$), and polymer-arginine interactions ($\Delta \Delta E_{pa}$). Increasing arginine concentration is denoted by increased shading (light to dark) and is indicated by arrows in (a-f). The polymer in water alone is shown in black, where appropriate. Mean values were estimated from three replicate REUS simulations. Error bars are reported as described in the SI. All plots are normalized to 0 at $R_g = 0.4$  nm, where appropriate.}
    \label{fig:decomp}
\end{figure*}

Fig.~\ref{fig:decomp}a shows the PMF along the $R_g$ reaction coordinate. In all solutions, free energy minima were observed at approximately 0.4 and 0.8 nm (configurations labeled I and III in Fig.~\ref{fig:decomp}g), along with a prominent free energy barrier at $\sim$0.6 nm representing the transition between folded and unfolded states (Fig.~\ref{fig:decomp}a). In pure water, hydrophobic collapse is unfavorable, with the unfolded state favored by $\sim$0.3 kT. In contrast, at all arginine concentrations, the folded state of the polymer is favored relative to pure water, and a monotonic increase in $\Delta G_{u}$ is observed (Fig. S4). An additional barrier at $\sim$0.45 nm was identified separating two folded states, labeled as I and II in Fig. \ref{fig:decomp}a.

The folded polymer ensemble in arginine solutions exhibits free energy minima corresponding to globular ($\sim$0.4 nm) and hairpin-like ($\sim$0.5 nm) configurations (labeled I and II in Fig.~\ref{fig:decomp}g). Arginine clusters encapsulating the hydrophobic polymer are observed in each state (Fig. S6). We propose that the free energy barrier separating these two states arises from an energetic penalty associated with breaking these encapsulating clusters. Such a mechanism is similar to that observed by Li et al.,\cite{li_solubilization_2010} who observed arginine-mediated suppression of hydrophobic association.

\subsection{Decomposition of arginine effects on polymer folding}
Fig.~\ref{fig:decomp}b-f shows the decomposition of the PMF into various components. The vacuum component, $W_{vac}$, favors the folded state of the polymer, associated with favorable intrapolymer interactions and configurational entropy upon collapse (Fig.~\ref{fig:decomp}b). Because this component does not depend on the presence of arginine, a balance of the remaining components dictates the effect of arginine on hydrophobic polymer collapse. 

Near large idealized solutes or hydrophobic interfaces, water dewets and forms a vapor-liquid-like interface.\cite{stillinger_structure_1973, lum_hydrophobicity_1999} Similar behavior may be expected for our hydrophobic polymer; hence, we computed $W_{cav}$, which depends on both the size and shape of the polymer and is related to the vapor-liquid surface tension.\cite{godawat_unfolding_2010, goel_attractions_2008, athawale_effects_2007} The cavitation component favors the folded state (Fig.~\ref{fig:decomp}c), reflecting a strong hydrophobic driving force for polymer collapse.\cite{ten_wolde_drying-induced_2002}

Attractive polymer-water interactions become more favorable with increasing $R_g$ (Fig.~\ref{fig:decomp}d), indicating polymer-water interactions oppose polymer collapse. It is worth noting that the free energy minima at $\sim$0.8 nm in the unfolded ensemble is observed in an aqueous environment (Fig.~\ref{fig:decomp}a), but not in vacuum (Fig.~\ref{fig:decomp}b). This minima arises due to favorable polymer-water interactions, consistent with prior MD simulations that showed water-mediated interactions drive large hydrophobic solutes apart.\cite{makowski_potential_2010,ben-amotz_water-mediated_2016,nayar_cosolvent_2018} Additionally, sufficient dewetting of the hydrophobic polymer is a known bottleneck to collapse,\cite{dhabal_characterizing_2021,ten_wolde_drying-induced_2002,athawale_effects_2007} resulting in the free energy barrier at $\sim$0.6 nm separating the folded and unfolded states. 

Attractive polymer-arginine interactions approach an energetic minima at $\sim$0.5 nm (Fig.~\ref{fig:decomp}e), giving rise to the global minimum observed in the overall PMF. Polymer-Cl$^{-}$ interactions were observed on the order of thermal fluctuations, consistent with previous observations that Cl$^{-}$ ions are depleted from the local domain of hydrophobic solutes.\cite{kalra_salting-and_2001, ghosh_salt-induced_2005}

Fig.~\ref{fig:decomp}h shows the change in each component upon unfolding in arginine solution relative to that observed in water. In Fig.~\ref{fig:decomp}h, the first $\Delta$ arises from the difference between folded and unfolded states (e.g., $\Delta E = \langle E_{u} \rangle - \langle E_{f} \rangle$), while the second $\Delta$ arises from the free energy difference between arginine solution ($\Delta E_{arg}$) and water ($\Delta E_{wat}$) (e.g., $\Delta\Delta E = \Delta E_{arg} - \Delta E_{wat}$). 
With increasing arginine concentration, we observed an increasingly favorable cavitation component for folding (Fig.~\ref{fig:decomp}h). Lin and Timasheff \cite{lin_role_1996} previously connected protein stability with cavity formation and vapor-liquid surface tension. They proposed that the expansion of a protein-containing cavity requires more energy in the presence of an additive that increases surface tension at the cavity-solution interface. Experimentally, increasing arginine concentration has been observed to increase the vapor-liquid surface tension of aqueous solutions,\cite{kita_contribution_1994} which may explain the dependence of $\Delta\Delta G_{cav}$ observed in the present study.


Looking at the trends in $\Delta \Delta E_{pw}$, we found that in 0.25 M arginine solutions, the polymer-water component favors polymer unfolding relative to in pure water (Fig.~\ref{fig:decomp}h). At 0.5 M and 1.0 M arginine concentrations, this component favors polymer collapse. With arginine present, the local domain of the polymer exhibits a reduction in the average number of water molecules (Fig. S10), indicating an effective expulsion of water. This, in turn, diminishes polymer-water interactions that resist polymer collapse.

The polymer-arginine contribution favors the folded polymer state at 0.25 M and 0.5 M arginine concentrations relative to in pure water (Fig.~\ref{fig:decomp}h). However, at 1.0 M arginine concentrations, polymer-arginine interactions promote polymer unfolding. Together, these results indicate that neither direct nor indirect mechanistic hypotheses alone can describe the effects of arginine on hydrophobic polymer folding.

\subsection{Mechanistic flip with increasing arginine concentration}
\begin{figure}[h]
 \centering
 \includegraphics[width=0.75\textwidth]{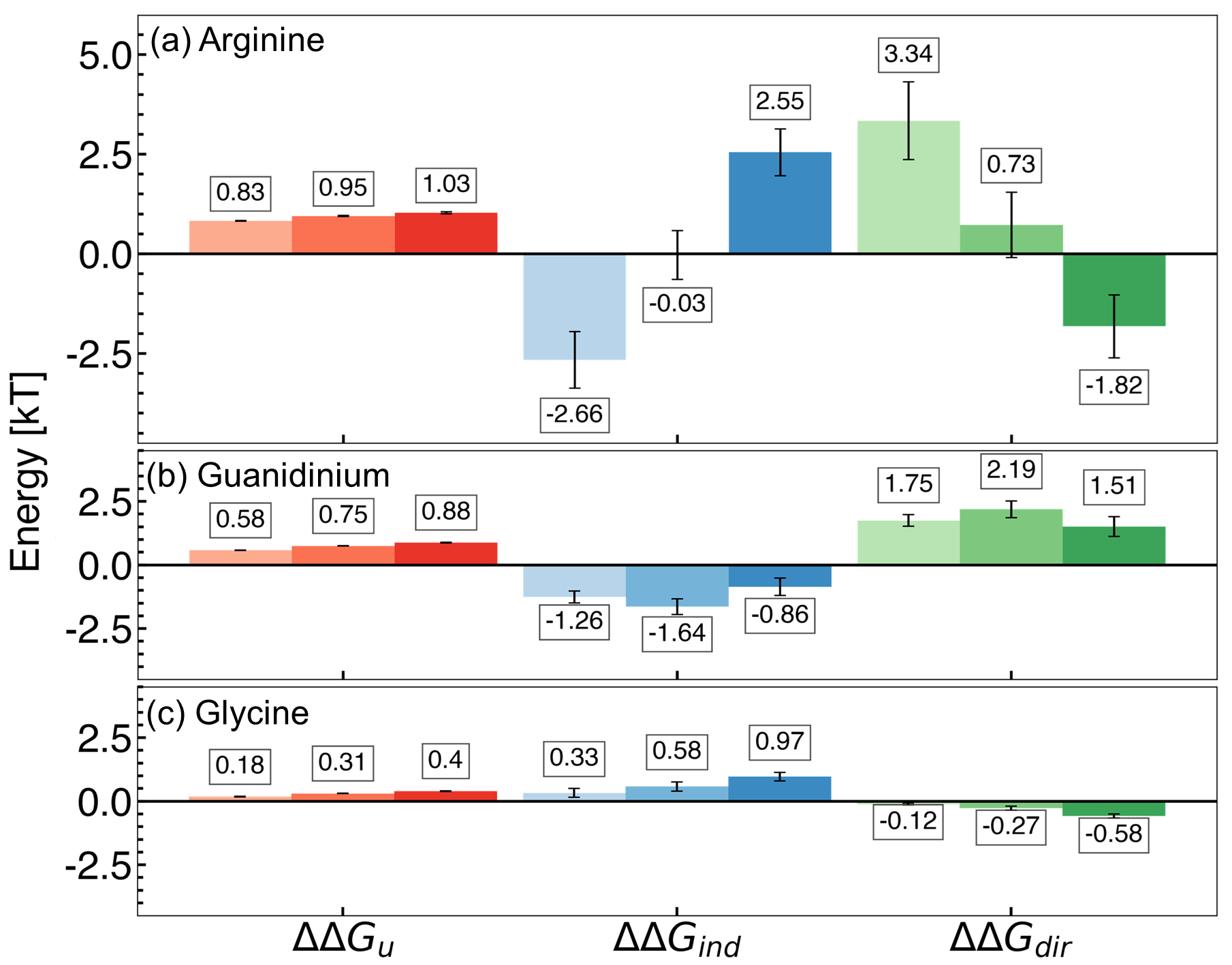}
 \caption{Contributions to the free energy of hydrophobic polymer unfolding in 0.25 M, 0.50 M, and 1.0 M (a) arginine, (b) guanidinium, and (c) glycine solutions. Changes in overall free energy of unfolding ($\Delta \Delta G_{u}$), direct interactions ($\Delta \Delta G_{dir}$), and indirect interactions ($\Delta \Delta G_{ind}$) and polymer are shown.  Increasing additive concentration is denoted by increased shading (light to dark; left to right). Mean values are reported from three replicate REUS simulations. Error bars were estimated via error propagation (see SI for details).}
    \label{fig:dbarsgg}
\end{figure}

To investigate potential competing effects of direct and indirect mechanisms, we combined the components of our PMF decomposition, delineating between those linked to direct effects (polymer-arginine and polymer-Cl$^{-}$; $\Delta\Delta G_{dir}$) and indirect effects (cavitation and polymer-water; $\Delta\Delta G_{ind}$) of arginine (Fig.~\ref{fig:dbarsgg}a).

We discovered that, with increasing concentration, the mechanism underlying the effects of arginine transitions from direct to indirect dominance. At 0.25 M, cavity formation and polymer-water interactions oppose polymer collapse, while arginine-polymer interactions favor collapse. The balance of these components gives rise to $\Delta\Delta G_{dir} > \Delta\Delta G_{ind}$, resulting in the net stabilization of folded conformations and supporting the direct mechanism hypothesis (Fig.~\ref{fig:dbarsgg}a).

In contrast, for the high-concentration regime (0.5 M and 1.0 M), cavity formation and polymer-water attractive interactions favor polymer collapse, while attractive arginine-polymer interactions favor extension of the hydrophobic polymer. In this case, indirect components dominate the free energy difference ($\Delta\Delta G_{dir} < \Delta\Delta G_{ind}$), stabilizing polymer collapse and supporting the indirect hypothesis (Fig.~\ref{fig:dbarsgg}a). 

Thus, within the range of concentrations studied, we have uncovered that arginine exists at the edge of a mechanistic flip between direct- and indirect-dominated stabilization of many-body hydrophobic interactions. The identification of this mechanistic switch may explain the wide variety of hypotheses in the existing arginine literature. Because arginine is situated on this razor's edge, small changes associated with the chemistry of a protein surface, the addition of cosolvents to solution, or differences in sample preparation, may cause significant changes in the modulation of hydrophobic interactions due to arginine. 

\subsection{Distinct roles of arginine's sidechain and backbone}
Arginine is comprised of a polar backbone and an aliphatic sidechain characterized by a guanidinium group. To investigate the roles of these components on hydrophobic polymer collapse, we completed an additional PMF decomposition in guanidinium and glycine solutions (see SI for simulation details). At all concentrations under study, we observed that guanidinium favors the hydrophobic polymer collapse primarily via a direct mechanism, while glycine stabilizes polymer collapse primarily via an indirect mechanism (Fig.~\ref{fig:dbarsgg}b,c).

In the case of guanidinium, stabilization is driven entirely by attractive polymer-guanidinium interactions that favor collapse, while polymer-water interactions and cavity formation oppose polymer folding (Fig.~\ref{fig:dbarsgg}b). In glycine solutions, however, stability is driven by the inverse mechanism; polymer-water interactions and the cavitation component favor collapse, while folding is opposed by attractive polymer-glycine interactions (Fig.~\ref{fig:dbarsgg}c). Based on these findings, we characterize arginine as exhibiting a guanidinium-like mechanism at low concentrations and a glycine-like mechanism at high concentrations.

While glycine is known to be an effective stabilizer of proteins,\cite{arakawa_stabilization_1985, bauer_impact_2017, platts_controlling_2015} our observations obtained for guanidinium are somewhat surprising due to its common role as a protein denaturant.\cite{watlaufer_nonpolar_1964, meuzelaar_guanidinium-induced_2015, jha_kinetic_2014} Several studies have stressed the importance of direct interactions in guanidinium-induced denaturation, primarily via breaking salt bridges, competing for intra-protein hydrogen bonds, and interacting with aromatic moieties via cation-pi stacking.\cite{meuzelaar_guanidinium-induced_2015, dempsey_reversal_2007, heyda_guanidinium_2017} Usually, this occurs at high concentrations of guanidinium salts. Our findings suggest that while guanidinium may stabilize hydrophobic interactions at low concentrations, this is outweighed by denaturing mechanisms at high concentrations.


\subsection{Resolving polymer-arginine-water interactions}
Thermodynamic analyses of arginine effects on polymer collapse discussed above indicate that direct and indirect mechanisms co-exist and compete at all concentrations under study. To probe this further, we characterize the molecular interactions between arginine, water, and the polymer. Specifically, we look at hydrogen bonding between arginine and water to characterize arginine-water interactions. Preferential interactions are used to elucidate the balance of polymer-arginine-water interactions. 

\begin{figure}[h]
   \includegraphics[width=0.75\textwidth]{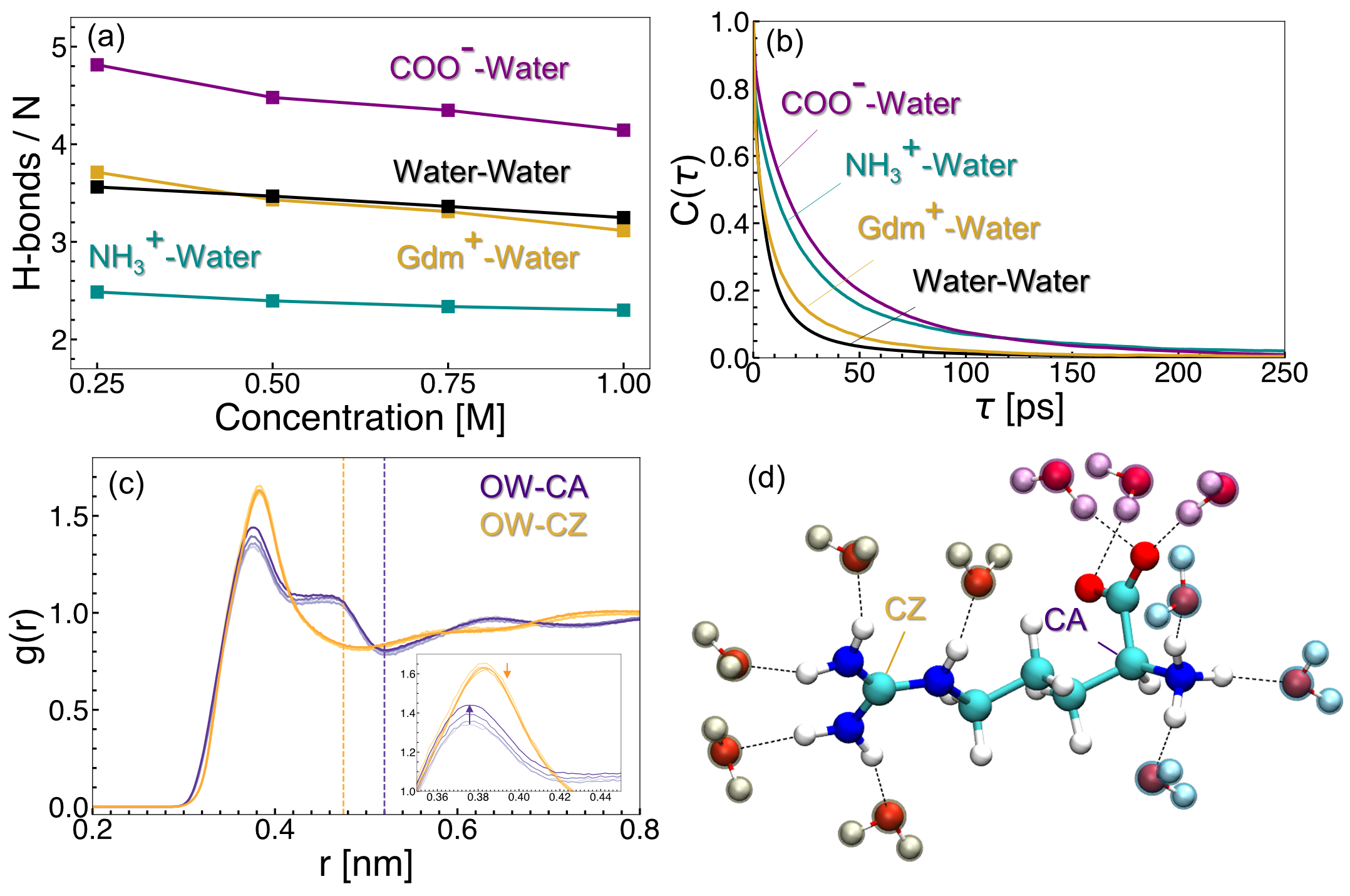}
    \caption{Arginine-water interactions. (a) Hydrogen bonds observed between backbone groups (COO$^{-}$, NH$_{3}^{+}$)  and the guanidinium (Gdm$^{+}$) sidechain with water. (b) Hydrogen bond existence correlation functions for water-water, COO$^{-}$-water, Gdm$^{+}$-water, and NH$_{3}^{+}$-water in 0.25 M arginine solution. (c) Radial distribution function between OW and either CA (purple) or CZ (gold). (c, inset) Arrows denote trends observed with increasing arginine concentration. (d) A representative snapshot of hydrogen-bonding interactions involving arginine and water. Water molecules interacting with the Gdm$^{+}$ sidechain are highlighted in yellow, while those interacting with NH$_{3}^{+}$ and COO$^{-}$ are shaded in blue and purple, respectively.}
    \label{fig:hbond}
\end{figure}

To describe how arginine interacts with water, we considered hydrogen bonding interactions between water and backbone (COO$^{-}$, NH$_{3}^{+}$) or sidechain (Gdm$^{+}$) atoms of arginine. Overall, the number of backbone-water hydrogen bonds was observed to be greater than sidechain-water hydrogen bonds (Fig.~\ref{fig:hbond}a). We further observed the fraction of occupied hydrogen bonding sites to be higher for backbone groups than the sidechain (Fig. S7). The hydrogen bond existence autocorrelation function for the 0.25 M arginine solution revealed hydrogen bonds formed between backbone-water atoms are, on average, longer-lived than sidechain-water hydrogen bonds (Fig.~\ref{fig:hbond}b). As arginine concentration was increased, hydrogen bond lifetimes were also observed to increase for both arginine-water and water-water interactions (Fig. S8). In principle, this may be considered a stabilizing property of arginine, as a growing body of literature has supported the role of stabilizing osmolytes in increasing hydrogen bond lifetimes and reducing water dynamics.\cite{hishida_effect_2022, jas_reorientation_2016, diaz_effect_2023, saladino_simple_2011, jahan_conformational_2020, zeman_effect_2020, gazi_conformational_2023}

Pairwise radial distribution functions (RDFs) were computed between the water oxygen (OW) and either the alpha carbon (CA) or guanidinium carbon (CZ) of arginine to quantify the local structure of water around arginine molecules (Fig.~\ref{fig:hbond}c). The first peak in the OW-CA RDF was observed to increase slightly with concentration. This indicates preferential hydration of the backbone group as more arginine molecules are introduced to the solution. There is, however, no such change observed in the OW-CZ RDF with concentration. A representative snapshot of water interactions with a single arginine molecule is shown in Fig.~\ref{fig:hbond}d (2D representation is shown in Fig. S9). Together, these results indicate that the backbone of arginine is the primary site for interaction with water. 

While we found arginine preferentially interacts with water via its backbone, we hypothesized arginine interacts with the polymer via its sidechain. It has been reported elsewhere that dehydration of the planar guanidinium face is important in forming face-face stacking interactions in aqueous guanidinium solutions.\cite{vazdar_arginine_2018} In our case, the dehydrated face of guanidinium is expected to play a key role in direct arginine-polymer interactions, similar to interactions observed between guanidinium and hydrophobic/aromatic protein residues.\cite{gund_guanidine_1972,tsumoto_role_2004,shukla_interaction_2010,arakawa_suppression_2007} 

\begin{figure*}[h]
   \includegraphics[width=1.0\textwidth]{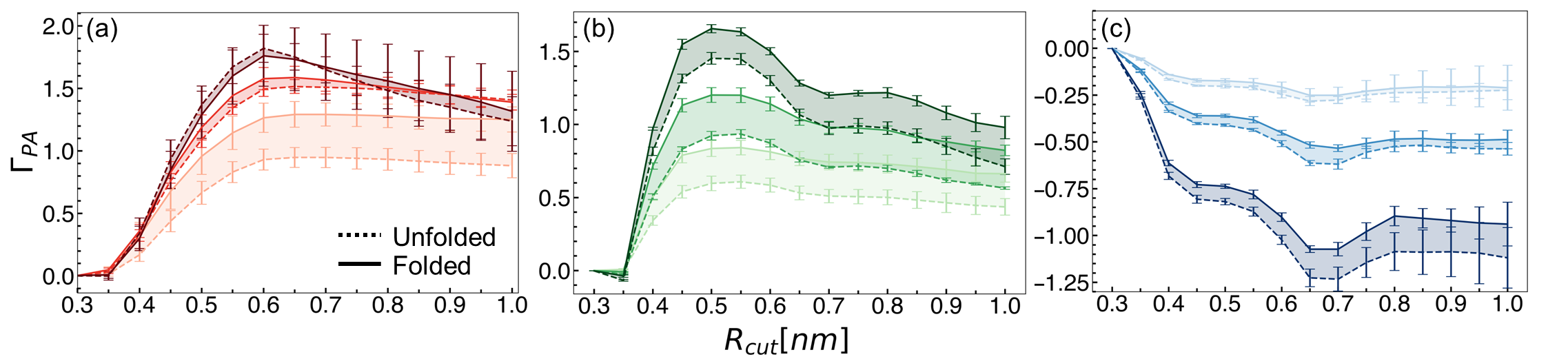}
    \caption{Preferential interaction coefficient values as a function of the cut-off distance for the local domain of the hydrophobic polymer for (a) arginine, (b) guanidinium, and (c) glycine. Dashed lines indicate values for the unfolded state, while solid lines indicate the folded state. Increasing concentration is denoted by increased shading (light to dark). Mean values and errors were estimated from three replicate simulations. Errors are reported as standard deviations from mean values.}
    \label{fig:prefint}
\end{figure*}

Wyman-Tanford theory relates the dependence of any equilibrium process (such as protein folding) and preferential interaction as:\cite{wyman_linked_1964, timasheff_protein-solvent_2002, shukla_molecular_2011}

\begin{equation}
    -\left(\frac{\partial\Delta G_{u}}{\partial\mu_{A}}\right) = \Gamma_{PA}^{u} - \Gamma_{PA}^{f}
\end{equation}
\noindent where $\Gamma_{PA}^{u}$ represents the preferential interaction coefficient in the unfolded state, while $\Gamma_{PA}^{f}$ represents the folded state. As a result, denaturants are expected to have a greater preferential interaction coefficient in the unfolded ensemble, while stabilizing osmolytes have a greater preferential interaction coefficient in the folded ensemble.\cite{canchi_cosolvent_2013, mondal_when_2013, mondal_how_2015, mukherjee_unifying_2020} In 0.25 M arginine concentration, we observed greater preferential interactions with the folded state relative to the unfolded state (Fig.~\ref{fig:prefint}a; Fig. S11a-c). With increasing concentration, we observe a diminishing difference between $\Gamma_{PA}^{u}$ and $\Gamma_{PA}^{f}$, in line with the mechanistic flip from a direct- to indirect-dominated stabilization mechanism.

Preferential interaction coefficients for guanidinium and glycine solutions are shown in Fig.~\ref{fig:prefint}b and c, respectively. We observed that at all concentrations, guanidinium preferentially interacts with the hydrophobic polymer (Fig.~\ref{fig:prefint}b). This is consistent with experimental evidence, as well as a prior simulation study that observed attractive guanidinium-polymer interactions with a model hydrophobic polymer.\cite{athawale_enthalpyentropy_2008, godawat_unfolding_2010} Glycine, meanwhile, was found to be preferentially excluded from the local domain of the hydrophobic polymer (Fig.~\ref{fig:prefint}c). This finding is consistent with the observed preference for the backbone of arginine to hydrogen bond with water, relative to the sidechain. Elsewhere, glycine has been observed to deplete from the surface of several model miniproteins, consistent with our findings. \cite{mukherjee_unifying_2020}

\begin{figure}[h]
   \includegraphics[width=0.75\textwidth]{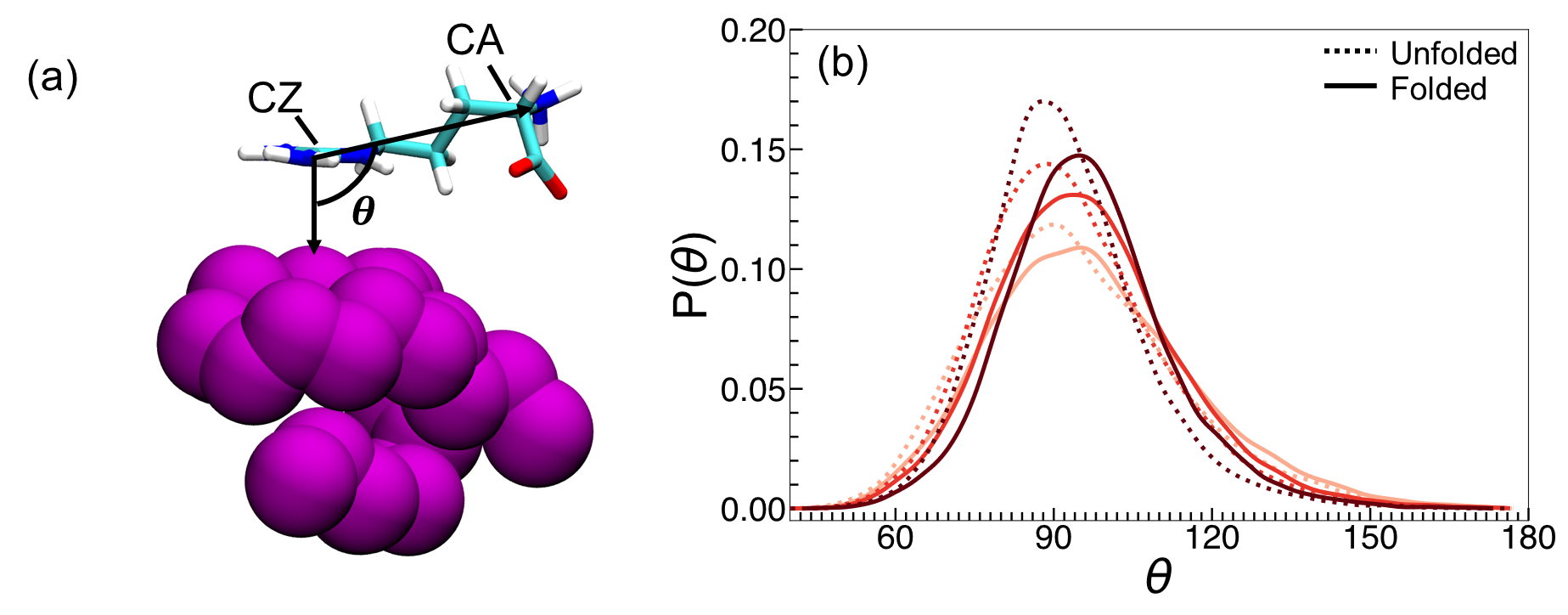}
    \caption{Preferential orientation of arginine relative to the hydrophobic polymer. (a) Representation of the three-body angle, $\theta$. (b) $P(\theta)$ is shown for 0.25 M, 0.5 M, and 1.0 M arginine concentrations. Solid lines denote the probability distribution for folded conformations, while dashed lines indicate the unfolded state. Increased concentration is denoted by increased shading.}
    \label{fig:orient}
\end{figure}

To explore whether the preferential interactions of arginine with polymer and water are accompanied by preferential orientations, we computed an orientation parameter inspired by Shukla and Trout.\cite{shukla_preferential_2011} This parameter was computed as the angle formed in three-dimensional space between polymer-CZ and CZ-CA vectors (Fig.~\ref{fig:orient}a). The monomer closest to CZ is taken for the polymer-CZ vector. Angles where $\theta > 90 \degree$ indicate the arginine backbone orients towards the bulk solvent, while $\theta < 90 \degree$ indicates the arginine backbone orients towards the polymer. We observed that for all concentrations, the probability $P(\theta)$ is skewed towards angles greater than 90$\degree$ (Fig.~\ref{fig:orient}b). 

Further, for all arginine concentrations, the mean value of $\theta$ for the folded state is greater than that observed for unfolded configurations (Fig.~\ref{fig:orient}b). This preferential orientation of arginine enables the hydrophobic face of the guanidinium sidechain to interact with the hydrophobic polymer, while extension of the backbone towards the bulk enables additional interactions with either water or other free arginine molecules. The greater ability for arginine to adopt preferred orientations in the folded state, particularly at 0.25 M, may partially explain the favorable $\Gamma_{PA}$ values described previously.

\section{Contextualizing the Mechanistic Flip of Arginine}
\subsection{Arginine destabilizes a charged polymer}
\begin{figure}[h]
   \includegraphics[width=0.55\textwidth]{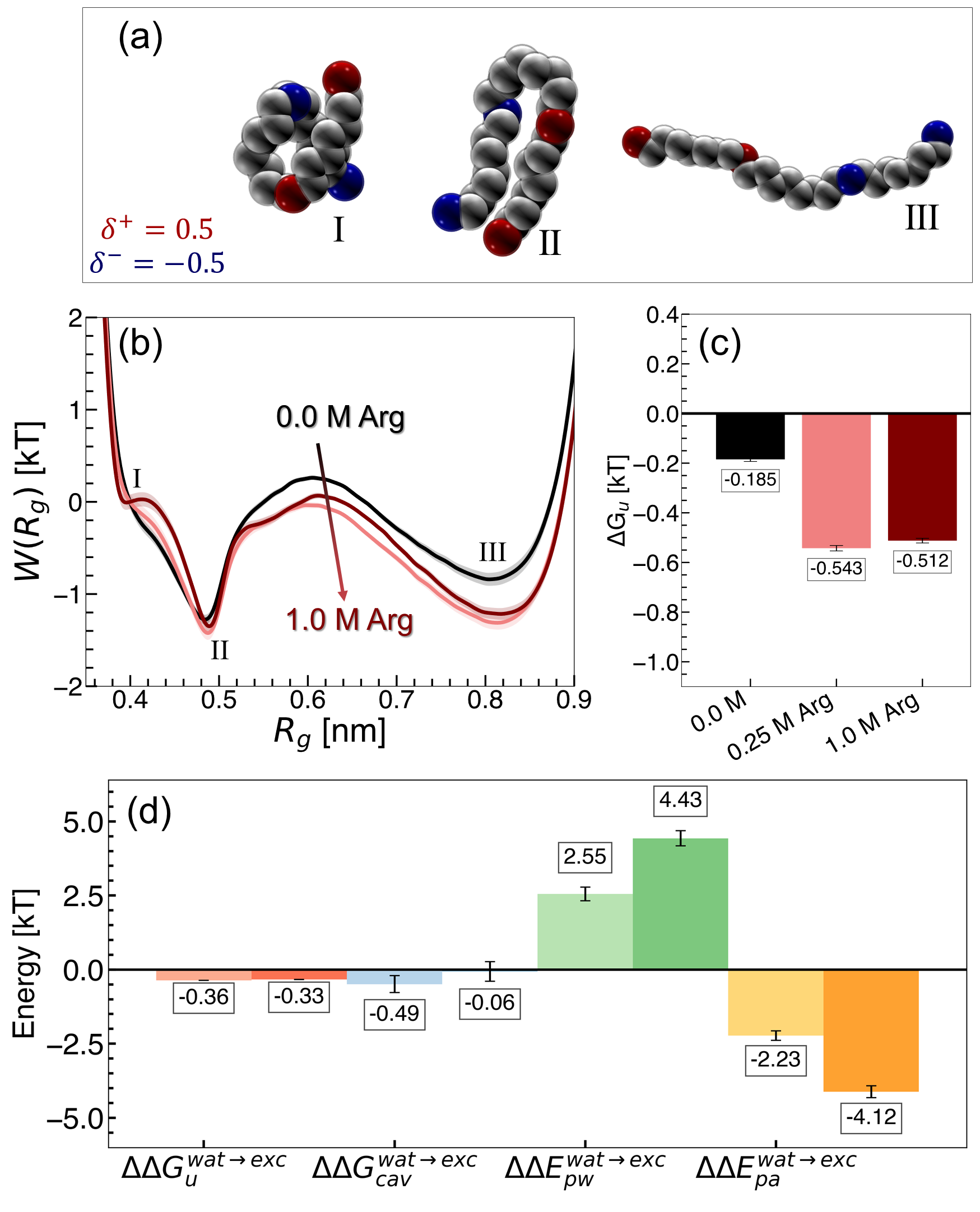}
    \caption{Charged polymer simulations. (a) Locations of the negative (blue) and positive (red) partial charges. (b) PMFs associated with the charged polymers. (c) Free energy of polymer collapse as a function of arginine concentration. (d) Change in PMF components in Arg solution relative to pure water, for the charged polymer. Increasing concentration is denoted by increased shading.}
    \label{fig:chargedPol}
\end{figure}

To explore whether we could shift arginine further towards a direct-dominated mechanism, we modified our hydrophobic polymer to include four beads with opposing partial charges (Fig. ~\ref{fig:chargedPol}a). In pure water, charged polymer collapse was observed to be favorable, resulting in a prominent free energy minimum at $R_{g} \sim$ 0.5 nm (Fig. ~\ref{fig:chargedPol}b). Relative to the hydrophobic polymer, this is indicative of a conformational preference of the polymer to adopt hairpin-like, rather than globule-like, configurations (I and II in Fig. ~\ref{fig:chargedPol}a, respectively).

Upon addition of either 0.25 M or 1.0 M arginine, folding of the charged polymer becomes less favorable (Fig. ~\ref{fig:chargedPol}c). We attribute this to a preference for arginine to interact with the charged sites, which are more accessible to arginine in the unfolded state. Indeed, at both concentrations, attractive polymer-arginine interactions dominate (Fig. ~\ref{fig:chargedPol}d), driving polymer unfolding. This model demonstrates that even subtle changes to the chemistry of a macromolecule can re-balance arginine mechanisms.

\subsection{Implications for formulation design}
Arginine-induced destabilization of a charged polymer illustrates that, when hydrophobic interactions compete with other effects, the effectiveness of arginine as a stabilizer can be altered. To better understand this feature of arginine, we explored the temperature stability of two models for formulation design: PPV and HEWL. 

PPV is a non-enveloped virus with a single stranded DNA genome.\cite{streck_porcine_2020} The PPV capsid is a spherical shell comprised of 60 copies of viral proteins (VP) VP1, VP2, and VP3 in a 1:10:1 ratio, arranged in an icosahedral symmetry.\cite{simpson_structure_2002} Due to the relatively small size and structural simplicity of PPV, recent studies have employed the virus as a model for investigating virus purification and thermostabilization techniques.\cite{mi_thermostabilization_2020, gencoglu_porcine_2014, tafur_reduction_2013, joshi_design_2024}

HEWL, meanwhile, is a relatively small protein with a well-defined fold (PDB: 1LYZ) and high stability.\cite{diamond_real-space_1974, matthews_studies_1995} Due to these features, it has been widely used as a model for protein folding\cite{dobson_understanding_1994, yoshimura_mechanism_1991, merlini_lysozyme_2005} and exploring osmolyte effects.\cite{santoro_increased_1992, adamczak_molecular_2016, arakawa_stabilization_1985} HEWL has also been used to investigate aggregation-suppressing effects of arginine. Several studies have proposed that arginine interacts favorably with aromatic and acidic residues of HEWL, which limits solvent exposure of aggregation-prone patches.\cite{tsumoto_role_2004, arakawa_suppression_2007, shukla_interaction_2010, vagenende_protein-associated_2013, brudar_effect_2023} Arginine has also been observed to enhance the heat-induced aggregation of bovine serum albumin and $\beta$-lactoglobulin, but not HEWL -- highlighting its context-dependent effects.\cite{shah_effects_2011}

\begin{figure}[h]
   \includegraphics[width=0.75\textwidth]{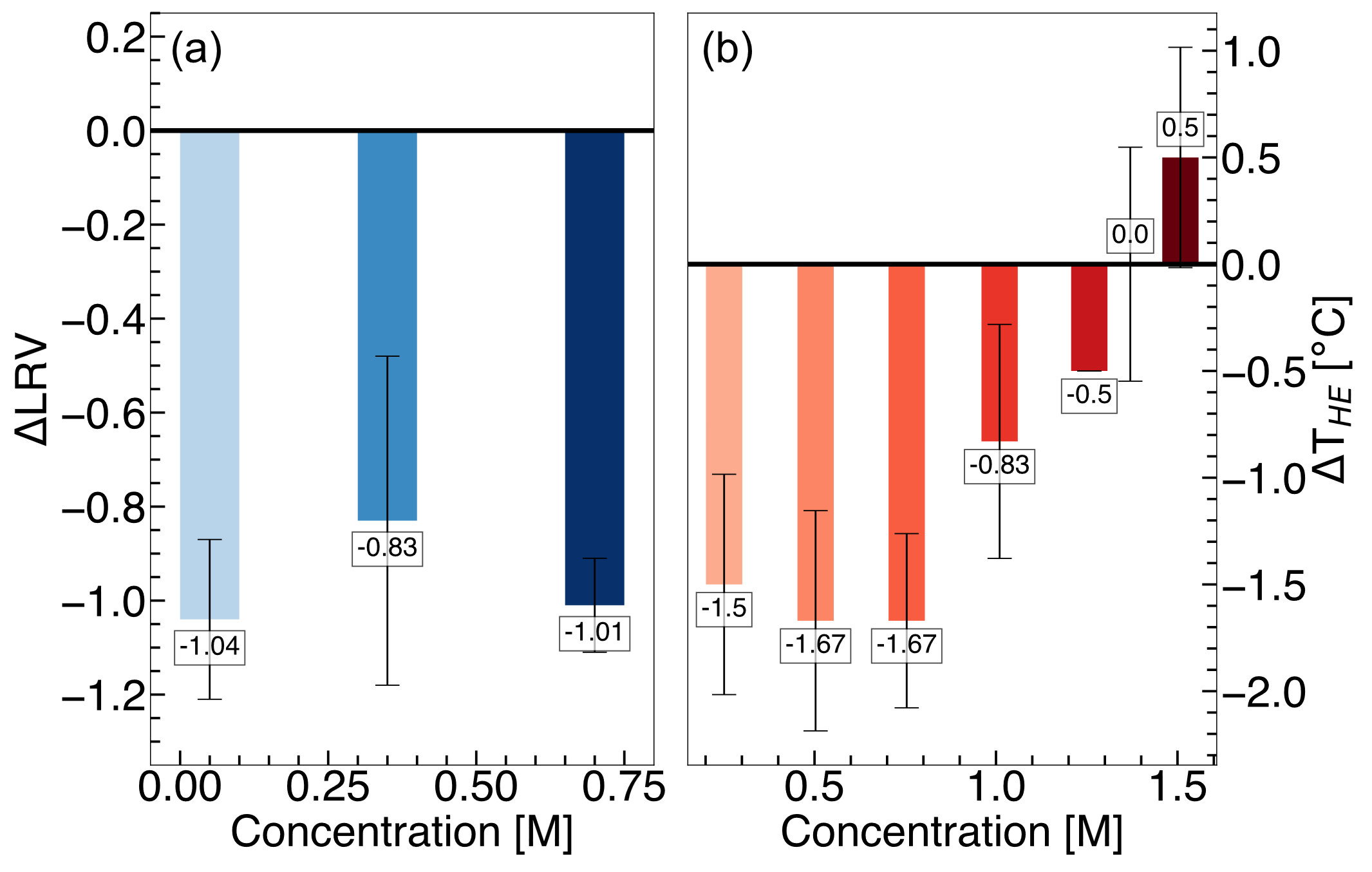}
    \caption{Temperature stability of (a) PPV and (b) HEWL as a function of arginine concentration. $\Delta$LRV is reported as LRV$_{Buffer}$ - LRV$_{Arg}$, while $\Delta$T$_{HE}$ $=$ T$_{HE, Arg}$ - T$_{HE, buffer}$. Increasing concentration is denoted by increased shading. Mean and standard deviation are estimated from triplicate measurements.}
    \label{fig:expt}
\end{figure}

To understand the effect of arginine on these biomolecules, temperature stability assays were carried out at different concentrations of arginine. For PPV, we completed an infectivity assay (see SI for experimental details) following virus incubation at 60 \degree C for 3 days -- sufficiently long to observe a significant decrease in the infectious titer.\cite{mi_thermostabilization_2020} Log reduction values (LRV) describe the decrease in the infectious titer of heat-treated virus, relative to the initial virus solution. We found, at all investigated arginine concentrations, more infectious PPV is lost relative to in buffer alone, resulting in negative $\Delta$LRV values (Fig. ~\ref{fig:expt}a). Such a finding indicates reduced temperature stability of the virus in arginine solutions.

For HEWL, we completed a thermal shift assay (see SI for details) to quantify the hydrophobic exposure temperature ($T_{HE}$) of the protein.\cite{seabrook_high-throughput_2013, shi_dsf_2013} At low concentrations (0.2-1.0 M) of arginine, a decrease in $T_{HE}$ is observed relative to buffer alone, indicating destabilization of HEWL (Fig. ~\ref{fig:expt}b). With increasing arginine concentration (1.5 M), the temperature stability of HEWL improves. Such a finding closely resembles the temperature stability of ovalbumin and lysozyme in arginine solutions reported by Vagenende \textit{et al.}\cite{vagenende_protein-associated_2013} Additionally, 1.0 M arginine was found to increase protein interactions with hydrophobic chromatographic materials.\cite{vagenende_protein-associated_2013} This finding highlights that, when experimentally-observed effects of arginine mirror what we observed with the hydrophobic polymer, hydrophobic interactions are likely to dominate. On the other hand, our results contrast findings presented by Platts and Falconer,\cite{platts_controlling_2015} who observed temperature-stabilizing effects at low arginine concentrations and destabilization of HEWL at high concentrations.

\subsection{Molecular-level investigation of formulation models}
The lack of consensus regarding the effects of arginine -- further accentuated by our investigation of PPV and HEWL temperature stability -- may be rationalized by the positioning of arginine at the edge of the mechanistic flip described in the present work. Additionally, we have shown that when charged beads are added to a hydrophobic polymer, this subtle change results in complete arginine-induced destabilization. To better understand the connection between the mechanistic flip of arginine and the relevance of differing molecular contexts, we completed straightforward MD simulations of HEWL and PPV systems in arginine solution.

The PPV major capsid protein, VP2, is known to self-assemble into virus-like particles that are non-pathogenic, but are morphologically similar to native PPV virions.\cite{rueda_effect_2000} Experimental identification of this structure\cite{simpson_structure_2002} enables PPV for molecular-level investigations via molecular simulations. On the capsid surface, prominent structural features include a canyon surrounding a pore at the 5-fold axis of symmetry, protruding spikes located at the 3-fold axis of symmetry, and a dimple on the 2-fold axis of symmetry.\cite{simpson_structure_2002} We found that the use of 15 VP2 proteins is the minimum system size required to capture these structural features, enabling simulations at timescales accessible to MD without sacrificing atomistic resolution. Despite this truncation of the viral capsid, our resulting system size was $\sim$750,000 atoms.

\begin{figure}[h]
   \includegraphics[width=0.75\textwidth]{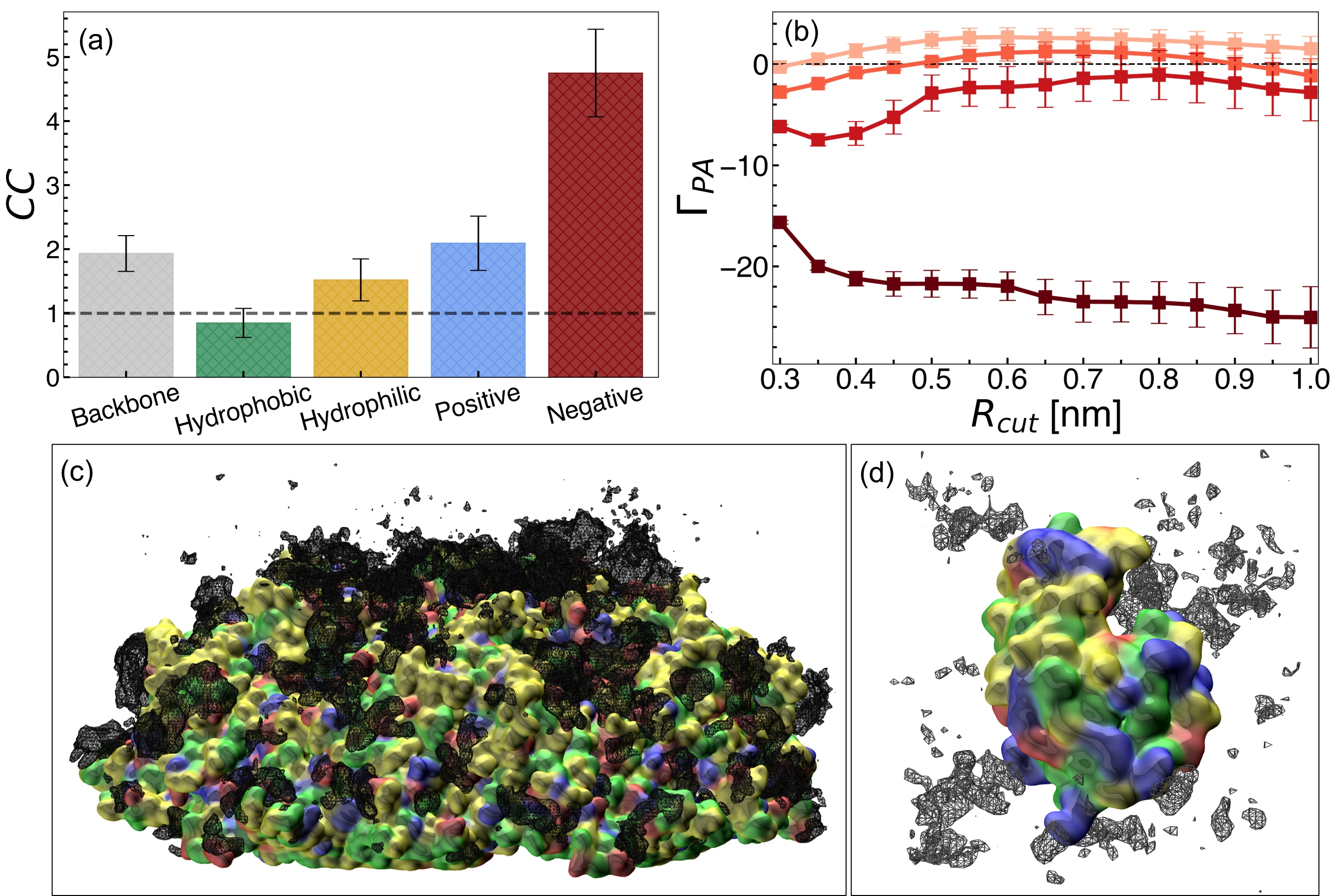}
    \caption{Arginine distributions from PPV and HEWL systems. (a) Arginine contact coefficients at the PPV 5-fold surface. Bars are shown for different residue types, hydrophobic (green), hydrophilic (yellow), positive (blue), and negative (red), and was computed separately for backbone atoms (gray). (b) Preferential interaction coefficient for arginine/HEWL. Increasing concentration is denoted by increased shading (light to dark). (c) Representative snapshot and volumetric density plots of the (c) PPV 5-fold surface and (d) HEWL in arginine solutions. In c and d, PPV and HEWL are shown in a surface representation. Protein residues are colored according to residue type as in panel a. Arginine density at an isovalue of 0.1 atoms / \AA$^{3}$ is shown in black mesh representation.}
    \label{fig:hewl-ppv}
\end{figure}

At the PPV surface in 0.25 M arginine solution, we observed preferential arginine accumulation at various sites across the surface (Fig. ~\ref{fig:hewl-ppv}a,c). Specifically, via calculation of contact coefficients of different residue types distributed across the PPV capsid surface, we identified significant accumulation of arginine near negatively charged glutamate and aspartate residues. From these findings, we hypothesize that direct interactions between arginine and charged residues at the PPV surface drive the instability observed from experiments. In this case, the dominating mechanism may be similar to that observed for arginine-induced unfolding of a charged polymer.

For HEWL, we completed simulations of 0.1 M, 0.2 M, 1.0 M, and 2.0 M arginine solutions. With increasing concentration, we observed an increase in preferential exclusion of arginine and reduction of arginine density local to the protein (Fig. ~\ref{fig:hewl-ppv}b,d). This finding, along with the increase in HEWL melting temperature observed at high arginine concentration, suggests that arginine imparts stabilization of HEWL via indirect effects. On the other hand, direct arginine-HEWL interactions may be destabilizing, which explains the decreased melting temperature of HEWL while arginine is less excluded from the local domain of the protein.

\section{Conclusions}
Overall, our findings illuminate the intricate mechanisms underlying the multi-faceted effects of arginine on hydrophobic polymer collapse. Arginine was observed to increase the favorability of many-body hydrophobic interactions, a key factor in protein stabilization. Our observations reveal a nuanced interplay in the impact of arginine on hydrophobic interactions, teetering on the edge of a mechanistic flip. At low concentrations, direct sidechain-driven interactions dominate, shifting to indirect backbone-driven effects at high concentrations.

The simultaneous presence of competing direct and indirect effects implies that changes in the chemistry of a protein surface, the addition of co-additives to solution, or differences in sample preparation may cause significant changes in the mechanism of action of arginine. A shift towards the direct mechanism risks guanidinium-like denaturation of native proteins by disrupting electrostatic and hydrogen-bonding interactions. Conversely, a shift towards the indirect mechanism may yield glycine-like stabilization of native proteins through preferential hydration. Indeed, we illustrated examples of the context-dependent effects of arginine through models of protein (HEWL) and virus (PPV) stability. We observed that arginine destabilizes PPV across a wide concentration range, which we attribute to destabilizing direct arginine-PPV interactions. This mechanism resembles the effects of arginine on a charged polymer model, suggesting that in both cases, arginine-induced destabilization via charge-charge interactions outweighs arginine-induced stabilization of hydrophobic interactions. We further found that arginine destabilizes HEWL at low concentrations and stabilizes HEWL at high concentrations, which we found is associated with an increased exclusion of arginine from the HEWL surface. This mechanism closely resembles the concentration-dependence of the indirect mechanism of arginine in a hydrophobic polymer model. Hence, at high concentrations, HEWL destabilization due to arginine charge-charge interactions may be opposed by an overall stabilization of hydrophobic interactions.

Together, our results suggest that arginine is uniquely situated for use in formulations due to its tunable, context-dependent properties. Hence, while arginine may not be considered the universal stabilizer it once was, its balance between direct- and indirect-driven stabilization of hydrophobic interactions solidifies its significance in formulation design.

\section*{Author Contributions}
J.Z.: Investigation - Computational, Validation, Formal Analysis, Visualization, Writing - Original Draft, Writing - Review and Editing; P.M.: Investigation - Computational, Validation, Writing - Review and Editing; I.T.: Conceptualization, Investigation - Experimental, Writing - Review and Editing; X.Z.: Investigation - Experimental, Writing - Review and Editing; C.H.: Conceptualization, Funding Acquisition, Resources, Writing - Review and Editing; S.P.: Conceptualization, Funding Acquisition, Resources, Writing - Review and Editing; S.S.: Conceptualization, Methodology, Validation, Supervision, Project Administration, Funding Acquisition, Resources, Writing - Review and Editing

\section*{Conflicts of interest}
There are no conflicts to declare.

\section*{Acknowledgements}
This material is based upon work supported by the National Science Foundation under DMREF Grant Nos. 2118788, 2118693, and 2118638.

\bibliography{ref}
\end{document}


\emergencystretch 3em

\section{Simulation Details}

\begin{figure}[!ht]
\centering
   \includegraphics[width=0.9\textwidth]{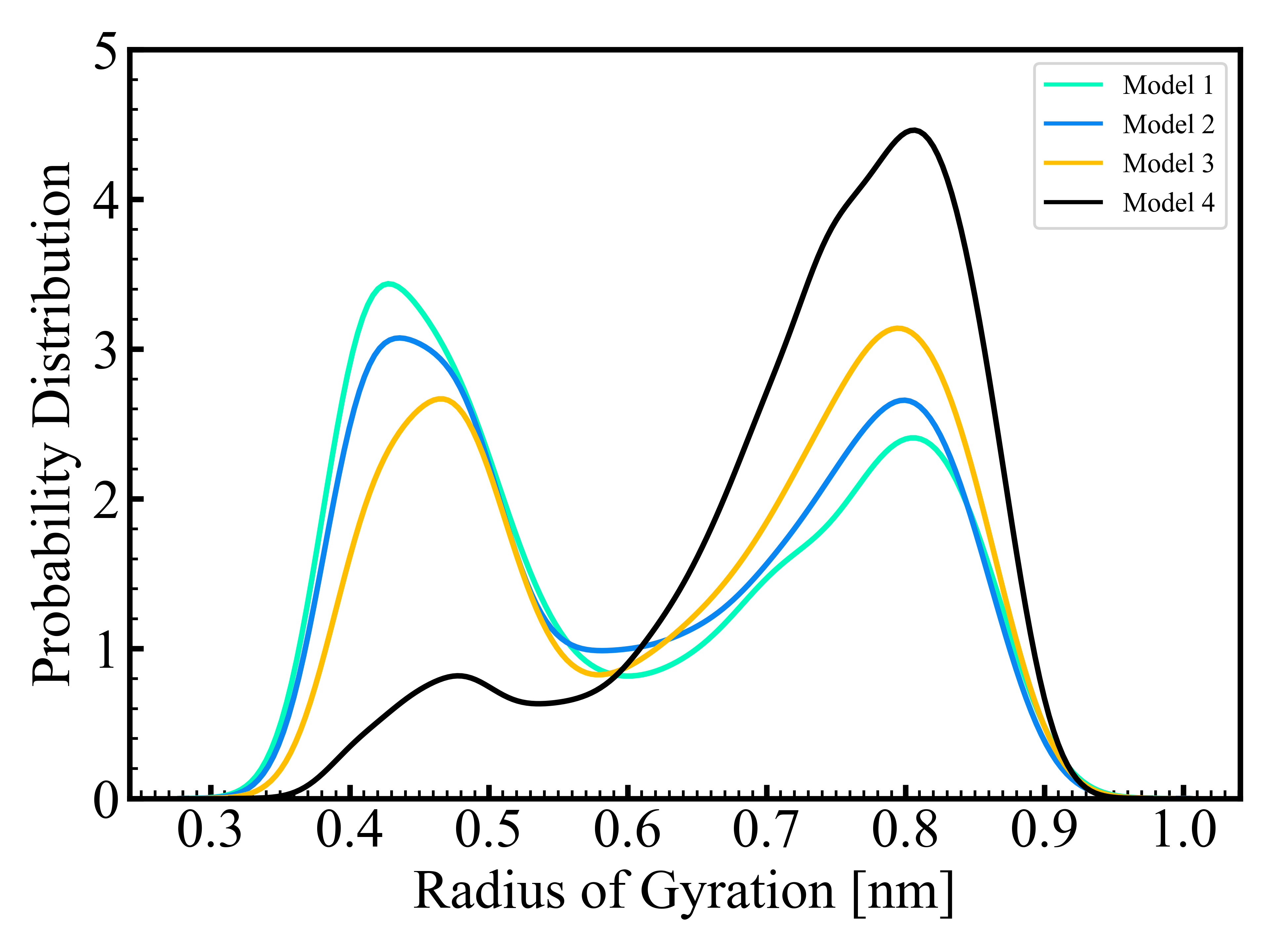}
   \caption{Probability distribution of radius of gyration obtained from 50 ns simulations of different polymer models in pure water. The models differ in their polymer-water interaction parameter, $\epsilon$, having 85\% (model 1), 88\% (model 2), 92\% (model 3), and 100\% (model 4) of the value calculated from Lorentz-Berthelot mixing rules.}
   \label{fig:si-poly-param-selection}
\end{figure}

A cubic box of length of 6.74 $\mathrm{nm}$ was constructed with a padding of 1.5 $\mathrm{nm}$ between the edge of the fully extended polymer and the nearest box edge. Chloride (Cl$^{-}$) counterions equal to the number of arginine molecules were added to achieve a net charge of zero. The TIP4P/2005\cite{abascal_general_2005} model was used for water, and the CHARM22 force field was used for arginine and Cl$^{-}$.\cite{brooks_charmm_2009} Lorentz-Berthelot mixing rules\cite{lorentz_1881, berthelot1898melange} were used to calculate non-bonded interactions between different atom types, except polymer-water oxygen interactions (Table \ref{tbl:si-polparam}). Polymer-water oxygen interactions were adjusted iteratively until the folded and unfolded states of the polymer were approximately evenly distributed in straightforward MD simulations. The various Lennard-Jones parameters tested are presented in Table \ref{tbl:si-polparam}. Guided by radius of gyration ($\mathrm{R_g}$) probability distributions, we selected parameters of model 2 for our study (Figure \ref{fig:si-poly-param-selection}).
\begin{table*}
  \caption{Polymer interaction parameters used in the present study.}
  \label{tbl:si-polparam}
  \begin{tabular}{c c c c}
    \hline
    Interaction & Model & Sigma (nm) & Epsilon (kJ/mol)\\
    \hline
    Polymer-Polymer & All& 0.373 & 0.586\\
    Polymer-Water & Model 1 & 0.345 & 0.573\\
    Polymer-Water & Model 2 & 0.345 & 0.593\\
    Polymer-Water & Model 3 & 0.345 & 0.620\\
    Polymer-Water & Model 4 & 0.345 & 0.674\\
    \hline
  \end{tabular}
\end{table*}

\begin{figure*}[!ht]
\centering
   \includegraphics[width=0.9\textwidth]{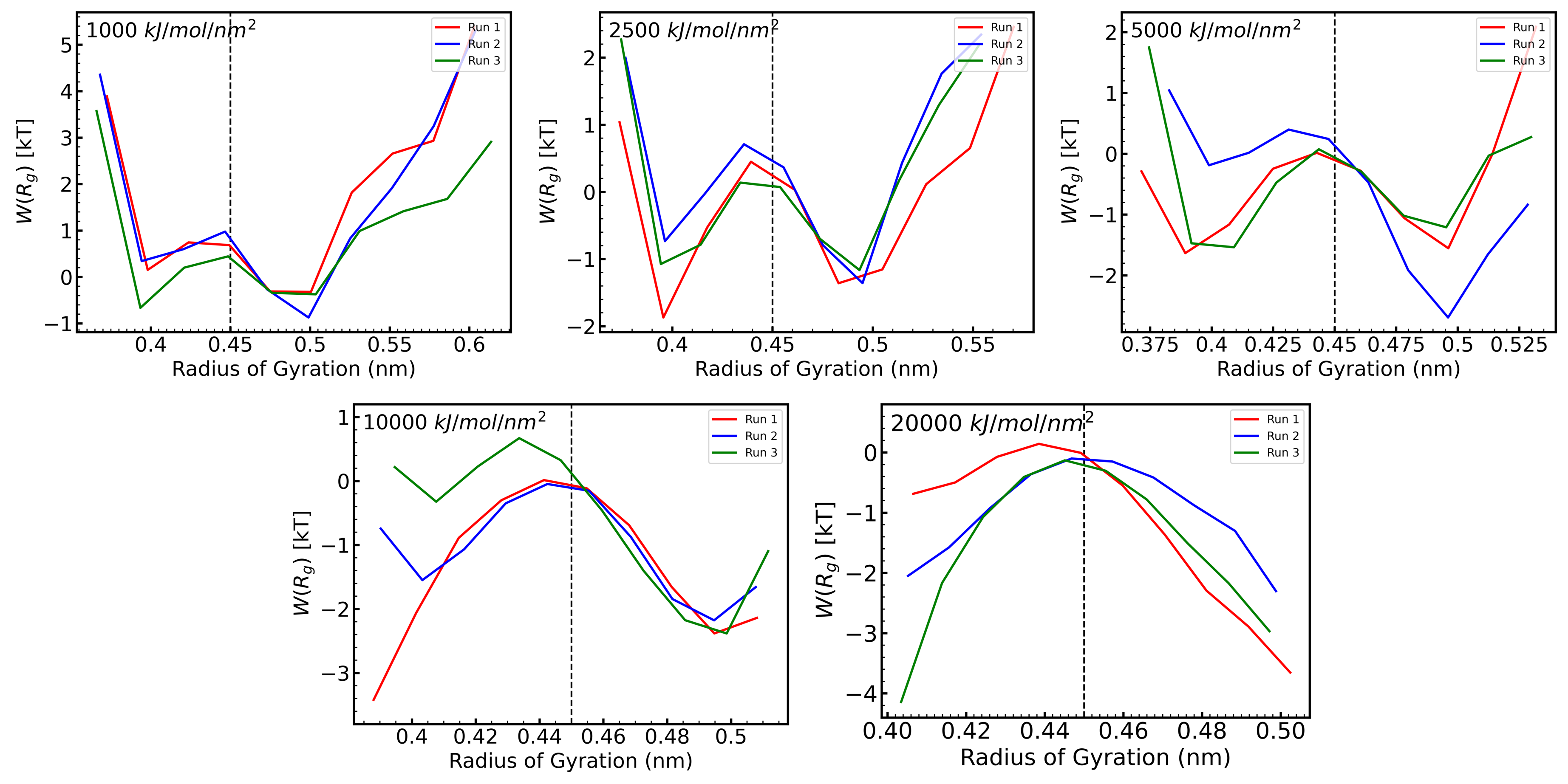}
   \caption{Sampling of 3 replicate runs in the window 3 (reference radius of gyration = 0.45 nm) region for polymer in 0.75M arginine solution with different force constants ranging from 1000 - 20000 $kJ/mol/nm^2$. For regions close to the reference, the uncertainty between runs is lesser for the lower force constant values. Force constant 1000 $kJ/mol/nm^2$ was chosen for window 3 based on these observations} 
    \label{fig:si-force-constant-sampling}
\end{figure*}

REUS simulations were performed in 12 evenly-spaced windows along the $R_g$ reaction coordinate, spanning 0.35 $\mathrm{nm}$ to 0.9 $\mathrm{nm}$. Each window was biased according to a harmonic potential, with a force constant of 1000 $\mathrm{kJ/mol/nm^2}$ for the window centered at 0.45 $\mathrm{nm}$ (window 3) and 5000 $\mathrm{kJ/mol/nm^2}$ for all other windows. We observed inefficient sampling in window 3 region (Fig. ~\ref{fig:si-force-constant-sampling}). Subsequent simulations with varying force constants for window 3 regions revealed that a force constant of 1000 $\mathrm{kJ/mol/nm^2}$ minimized differences between replicate runs in the regions close to the window centers. 

The windows are first energy minimized using the steepest descent minimization with a tolerance of 10 $\mathrm{kJ/mol/nm}$ and step size of 0.01. For each window, 1 $\mathrm{ns}$ NVT equilibration is then performed using V-rescale thermostat (temperature coupling time constant, $\mathrm{\tau_{T}}$ = 0.5 $\mathrm{ps}$),\cite{bussi_canonical_2007} followed by 1 $\mathrm{ns}$ NPT equilibration using the V-rescale thermostat ($\mathrm{\tau_T}$ = 0.5 $\mathrm{ps}$)\cite{bussi_canonical_2007}  and Berendsen barostat ($\mathrm{\tau_{P}}$ = 0.5 $\mathrm{ps}$)\cite{berendsen_molecular_1984} to bring the system to a temperature of 300 $\mathrm{K}$ and pressure of 1 $\mathrm{atm}$. NPT production run for 100 $\mathrm{ns}$ is simulated for each window using Nos\'e-Hoover temperature coupling ($\mathrm{\tau_{T}}$ = 5 $\mathrm{ps}$)\cite{evans_nosehoover_1985} and Parrinello-Rahman pressure coupling ($\mathrm{\tau_P}$ = 25 $\mathrm{ps}$).\cite{parrinello_polymorphic_1981} A Hamiltonian exchange move is attempted every 200 timesteps, with a 2 $\mathrm{fs}$ time step. The Particle Mesh Ewald (PME) algorithm was used for electrostatic interactions with a cut-off of 1 $\mathrm{nm}$. A reciprocal grid of 42 x 42 x 42 cells was used with $\mathrm{4^{th}}$ order B-spline interpolation. A single cut-off of 1 $\mathrm{nm}$ was used for van der Waals interactions. The neighbor search was performed every 10 steps.

To further investigate the hypothesis that attractive polymer-arginine interactions are driven by the guanidinium sidechain while indirect effects are driven by the backbone, additional independent REUS simulations including either guanidinium or glycine as the additive were carried out. Guanidinium parameters were based on the CHARMM22 parameters of arginine. This was achieved by truncating an arginine molecule up to the first guanidinium nitrogen, protonating this atom, and imposing a symmetric charge distribution according to the existing parameters. Glycine parameters were taken directly from the CHARMM22 force field. Systems in the same concentration range as arginine were generated to study sidechain and backbone contributions to hydrophobic polymer collapse.

\begin{table*}
  \caption{Setup of simulated systems.}
  \label{tbl:si-simsetup}
  \begin{tabular}{c c c c c}
    \hline
    System & Simulation Time (ns) & Concentration (M) & $N_{Exc}$ & $N_{Wat}$ \\
    \hline
    Arginine & 20 & 0.25 & 47 & 9653 \\
    Arginine & 20 & 0.50 & 93 & 9111 \\
    Arginine & 20 & 0.75 & 139 & 8582 \\
    Arginine & 20 & 1.0 & 185 & 7933 \\
    Polymer & 3 x 100 x 12 & 0.00 & 0 & 10599 \\
    Polymer + Arginine & 3 x 100 x 12 & 0.25 & 47 & 10092 \\
    Polymer + Arginine & 3 x 100 x 12 & 0.50 & 93 & 9511 \\
    Polymer + Arginine & 3 x 250 x 12 & 1.0 & 185 & 8398 \\
    Polymer + Guanidinium & 3 x 50 x 12 & 0.25 & 47 & 10364 \\
    Polymer + Guanidinium & 3 x 50 x 12 & 0.50 & 93 & 10144 \\
    Polymer + Guanidinium & 3 x 50 x 12 & 1.0 & 185 & 9702 \\
    Polymer + Glycine & 3 x 50 x 12 & 0.25 & 47 & 10318 \\
    Polymer + Glycine & 3 x 50 x 12 & 0.50 & 93 & 10022 \\
    Polymer + Glycine & 3 x 50 x 12 & 1.0 & 185 & 9444 \\
    \hline
  \end{tabular}
\end{table*}

\section{Error Calculations}

The errors for PMF were calculated through the propagation of uncertainty using 3 replicate simulations ($\mathrm{N=3}$). The derivation of uncertainty in the free energy of unfolding is shown below. $\mathrm{\sigma}$ represents the standard deviation, $\mathrm{exp}$ represents the exponential term, $\mathrm{ln}$ represents the logarithmic term and $\mathrm{int}$ represents the integral.  

\begin{equation}
    \Delta G_{\text{unfold}}=k_BT \ln \frac{\int_{R_g^{\text{cut}}}^{R_g^{\max }} \exp \left(-\frac{W\left(R_g\right)}{k_B T}\right) d R_g}{\int_{R_g^{\min }}^{R_g^{\text {cut}}} \exp \left(-\frac{W\left(R_g\right)}{k_B T}\right) d R_g}
\end{equation}

The integral is approximated as a sum and divided into discrete bins in the Rg coordinate. The Rg space (from 0.3 to 0.9 $\text{nm}$) is divided into 600 bins, giving a $\mathrm{\Delta R_g}$ = 0.001 $\text{nm}$.

\begin{equation}
    \sigma_{W\left(R_g\right)}=\sqrt{\frac{\sum\left(W\left(R_g\right)_i-\mu_{W\left(R_g\right)}\right)^2}{N}}
\end{equation}

\begin{equation}
    \sigma_{\text {exp }}=\left|\exp \left(-\frac{W\left(R_g\right)}{k_B T}\right)\right| *\left|\frac{1}{k_B T} * \sigma_{W\left(R_g\right)}\right|
\end{equation}

\begin{equation}
    \sigma_{\text {int}}= \Delta R_g * \sqrt{\sum \sigma_{\text{exp}}^2}
\end{equation}

\begin{equation}
    \sigma_{ln}=\frac{\sigma_{\text{int}}}{\text{int}}
\end{equation}

\begin{equation}
    \sigma_{\Delta G}=k_B T * \sqrt{\left(\sigma_{ln}\right)_{num}^2+\left(\sigma_{ln}\right)_{den}^2}
\end{equation}

The errors in PMF decomposition were calculated using error propagation rules. An example of error calculation for $\mathrm{\Delta E_{unfold}}$ is shown below:

\begin{equation}
    \Delta E_{unfold} = \langle E\rangle_u - \langle E\rangle_f
\end{equation}

\begin{equation}
    \langle E\rangle_f=\frac{\sum_{r_{\text {min }}}^{r_{\text {cut }}} E\left(R_g\right) P\left(R_g\right)}{\sum_{r_{\text {min }}}^{r_{\text {cut }}} P\left(R_g\right)}, \quad\langle E\rangle_u=\frac{\sum_{r_{\text {cut }}}^{r_{\text {max }}} E\left(R_g\right) P\left(R_g\right)}{\sum_{r_{\text {cut }}}^{r_{\text {max }}} P\left(R_g\right)}
\end{equation}

\begin{equation}
    \sigma_{E\left(R_g\right)}=\sqrt{\frac{\sum\left(E\left(R_g\right)_i-\mu_{E\left(R_g\right)}\right)^2}{N}}
\end{equation}

\begin{equation}
    \sigma_{\langle E\rangle}=\frac{\sqrt{\sum_{r_{\text {min }}}^{r_{\text {cut }}} \sigma_{E\left(R_g\right)}^2 P\left(R_g\right)^2}}{\sum_{r_{\text {min }}}^{r_{\text {cut }}} P\left(R_g\right)}
\end{equation}

\begin{equation}
    \sigma_{\Delta E}=\sqrt{\sigma_{int, f}^2 + \sigma_{int, u}^2}
\end{equation}

\section{Clustering Analysis}
\begin{figure*}[!ht]
\centering
   \includegraphics[width=0.9\textwidth]{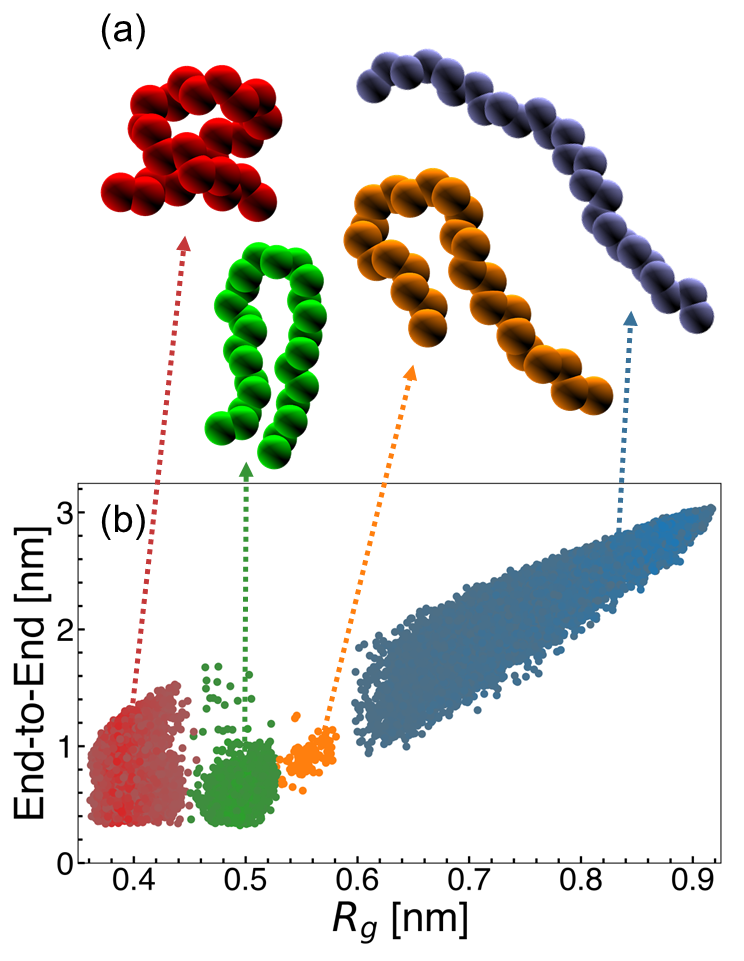}
    \caption{(a) Representative configurations from HDBSCAN clustering in 0.25 M arginine solution. (b) Polymer configurations projected onto end-to-end distance and radius of gyration space.}
    \label{fig:si-polclust}
\end{figure*}

Clustering was achieved via the leaf algorithm of HDBSCAN\cite{mcinnes_hdbscan_2017}. The minimum cluster size parameter was set to 100, while the minimum samples parameter was set to 50. Clustering was carried out on the principal moments of the gyration tensor of the hydrophobic polymer. Data were obtained from the final 100 ns in each window (3.6 $\mu$s total), saving coordinates every 100 ps. Data points not belonging to clusters were removed, for clarity. Clusters identified in principal moment space were projected onto end-to-end vs radius of gyration space. Representative snapshots are shown in Fig ~\ref{fig:si-polclust} to illustrate the configurations obtained in each cluster. Clusters at $R_g = 0.4$ and $R_g = 0.5$ are separated by a free energy barrier in the calculated PMFs.

\section{Preferential Interaction Coefficients}
\begin{figure*}[!ht]
\centering
   \includegraphics[width=1.0\textwidth]{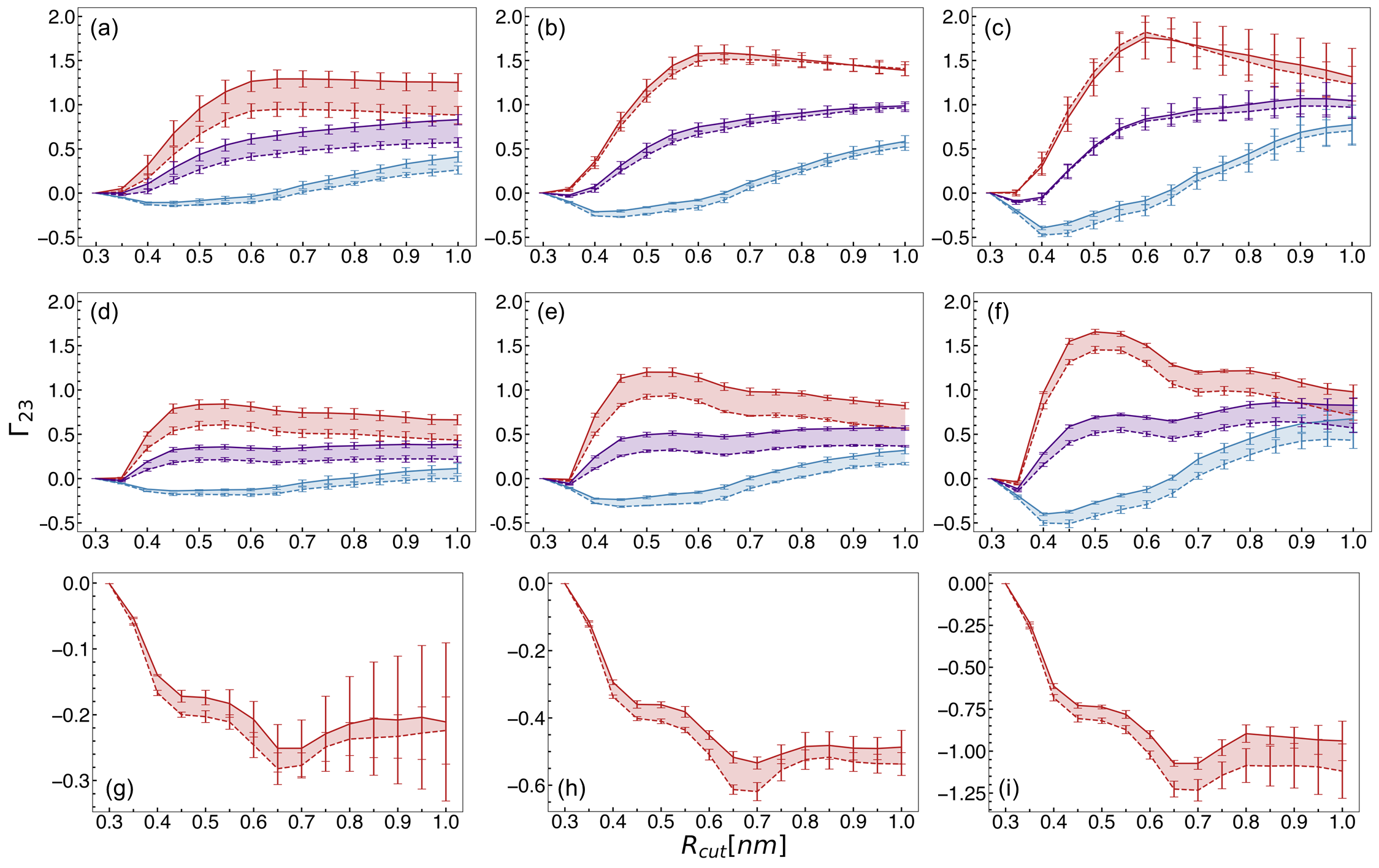}
   \caption{Preferential interaction coefficients for (a-c) arginine, (d-f) guanidinium, (g-i) glycine solutions. The additive is colored in red, counterion (if present) is colored in blue, and the net preferential interaction coefficient is colored in purple. Dashed lines indicate values for the unfolded state, while solid lines denote the folded state. Increasing arginine concentration is denoted by increased shading (light to dark). Mean values are reported from three replicate REUS simulations. Error bars were estimated as standard deviations from three replicate simulations.} 
    \label{fig:si-prefint}
\end{figure*}

In the main text, we denote water, polymer, and additive as W, P, and A, respectively. Here, we follow traditional notation found in literature, denoting water, polymer, and additive as 1, 2, and 3, respectively. At higher concentrations, no preference for folded versus unfolded conformations was observed. Cl$^-$ was found to preferentially deplete from the local domain of the polymer at both high and low concentrations (Fig. ~\ref{fig:si-prefint}), as expected. For a binary electrolyte such as ArgCl, the net preferential interaction coefficient is obtained as\cite{record_interpretation_1995}
\begin{equation}
    \Gamma_{23}=0.5(\Gamma_{23}^{-} + \Gamma_{23}^{+} - |Z|)
\end{equation}
where $\Gamma_{23,-}$ denotes the preferential interaction coefficient for the anion, $\Gamma_{23,+}$ for the cation, and $Z$ is the charge of the solute (for the polymer, $Z=0$). 

The net preferential interaction coefficient of the binary electrolyte ArgCl is reported in Fig.~\ref{fig:si-prefint}. The observed increase in $\Gamma_{23}^{ArgCl}$ with increasing concentration is in contrast to experimental evidence suggesting arginine tends to preferentially interact with proteins at low concentrations and becomes excluded with increasing concentration.\cite{shukla_interaction_2010, schneider_investigation_2009, schneider_arginine_2011, vagenende_protein-associated_2013} Our findings suggest that this concentration-dependent behavior of arginine is likely not mediated by the presence of hydrophobic interaction sites.

\section{Experimental Details}
\subsection{Temperature Stability of PPV}

\noindent\textbf{Materials}

Eagle’s minimum essential media (EMEM), sodium bicarbonate (7.5\% solution), penicillin/streptomycin (pen/strep, 10,000 U/ml), fetal bovine serum (FBS, qualified, USDA-approved regions), phosphate-buffered saline (1 X PBS, pH 7.2), and trypsin/EDTA (0.25\%) used for cell culture were purchased from Gibco™ (Grand Island, NY). MTT (2-(3,5-diphenyltetrazol-2-ium-2-yl)- 4,5-dimethyl-1,3-thiazole; bromide, 98\%) and sodium dodecyl sulfate (SDS, BioReagent, $\geq$98.5\%) were purchased from Fisher Scientific (Waltham, MA) for virus titration. Arginine monohydrochloride (reagent grade, $\geq$98\% (HPLC)) was purchased from Millipore Sigma (Burlington, MA) as the stabilizing excipient. Sodium phosphate monobasic monohydrate (reagent ACS grade) was purchased from Millipore. Sodium phosphate dibasic heptahydrate (ACS reagent, $\geq$98.0\%) was purchased from Sigma Aldrich (St. Luis, MO). 

\noindent\textbf{Methods}

\noindent\textit{Cell line and virus}

Porcine kidney cells (PK-13) were purchased from the American Type Culture Collection (ATCC\textsuperscript{\textregistered}) (cat\# CRL-6489™) and cultured in EMEM supplemented with 10 v/v\% FBS and 1 v/v\% pen/strep. The cells were incubated at 37 \degree C, 5\% CO2, and 100\% relative humidity. Porcine parvovirus (PPV) strain NADL-2 was a generous gift from Dr. Ruben Carbonell at North Carolina State University (Raleigh, NC). PPV strain NADL-2 was propagated in PK-13 cells using a previously established method \cite{heldt_colorimetric_2006}. After three freeze-thaw cycles, the cell lysate was clarified by centrifugation at 5,000 rpm at 4 \degree C for 15 minutes in an ST16R centrifuge with a TX-400 swing-bucket rotor (Thermo Scientific (Waltham, Ma)). The PPV-containing supernatant was stored at -80 \degree C prior to use.

\noindent\textit{Virus quantification} 

The titer of PPV was found by the MTT colorimetric cell viability assay \cite{joshi_design_2024}. PK-13 cells were seeded at a density of 8 × 10$^{4}$ cells/mL in 96-well plates and incubated overnight. The next day, the cells were infected with a 1:5 serial dilution of samples. After six days, 5 mg/mL of MTT in 1X pH 7.2 PBS was added to each well. Four hours later, 10 w/v\% SDS with 0.01 M hydrochloric acid (HCl) was added to each well and the absorbance at 550 nm was measured with a Synergy$^{TM}$ Mx microplate reader from BioTek (Winoski, VT) the next day. The 50\% viral infectious dose was determined in units of MTT$_{50}$/mL. 

\noindent\textit{Liquid viral sample preparation}

The excipient solutions were made by dissolving different concentrations of arginine monohydrochloride in phosphate buffer containing 1.54 mM sodium phosphate monobasic monohydrate and 2.71 mM sodium phosphate dibasic. The virus samples were made by adding 10 v/v\% viral stock solutions to the excipient solution. 

\noindent\textit{Thermostability studies}

Liquid samples were prepared in triplicates and were put either in a heat block at 60 \degree C \cite{mi_thermostabilization_2020} or in a fridge at 4 \degree C as the control samples. 72 hours later, the titer of virus in each sample was determined using the MTT assay.  

\subsection{Temperature Stability of HEWL}

\noindent\textbf{Materials}

Hen egg white lysozyme (HEWL $\geq$ 95\%) was purchased from Hampton Research. 4-(2-hydroxyethyl)-1-piperazineethanesulfonic acid) (HEPES $\geq$ 99\%), hydrochloric acid (HCl, ACS grade), and sodium hydroxide (NaOH, ACS grade) were purchased from Fisher Scientific. L-arginine hydrochloride ($\geq$ 99\%) was purchased from Sigma Aldrich. The Protein Thermal Shift$^{TM}$ Dye Kits) were purchased from Applied Biosystems.

\noindent\textbf{Methods}

\noindent\textit{Stock Solution Preparation}

Stock solutions of 1 M NaOH and 1 M HCl were prepared gravimetrically in DI water. A stock solution of 10 mM HEPES was prepared gravimetrically in DI water and adjusted to pH = 7.00 ± 0.03 with HCl and NaOH, as needed (Thermo Scientific ROSS Sure-Flow Combination pH). Stock solutions of 1.24 mM HEWL (18 mg/mL) and 1.64 M arginine were prepared in 10 mM HEPES. A stock solution of 50X Sypro Orange was prepared using dye and buffer provided in the Protein Thermal Shift kit.

\noindent\textit{Test Sample Preparation}

Samples were prepared as in Table ~\ref{tbl:si-samples} by mixing 10 mM HEPES, arginine, HEWL, and Sypro Orange to a microcentrifuge tube. Samples were mixed after the addition of HEWL and Sypro Orange via vortexing.

\begin{table*}
  \caption{Setup of simulated systems.}
  \label{tbl:si-samples}
  \begin{tabular}{c c c c}
    \hline
    50x Sypro Orange ($\mu$L) & 1.24 mM HEWL ($\mu$L) & 1.64 M ArgHCl ($\mu$L) & 10 mM HEPES ($\mu$L) \\
    \hline
    8 & 6 & 0 & 66 \\
    8 & 6 & 11 & 55 \\
    8 & 6 & 22 & 44 \\
    8 & 6 & 33 & 33 \\
    8 & 6 & 44 & 22 \\
    8 & 6 & 55 & 11 \\
    8 & 6 & 60 & 6 \\
    8 & 6 & 63 & 3 \\
    8 & 6 & 66 & 0 \\    
    \hline
  \end{tabular}
\end{table*}

\noindent\textit{Thermal Shift Characterization}

Differential scanning fluorimetry (DSF) can be used to determine the hydrophobic exposure temperature ($T_{HE}$) in a high-throughput manner.\cite{seabrook_high-throughput_2013, shi_dsf_2013} The Sypro Orange dye shows enhanced fluorescence upon binding to hydrophobic regions of a protein, allowing for detection of protein unfolding events as a function of temperature. Previous reports have shown a strong correlation between $T_{HE}$ and the actual thermodynamic melting temperature of a protein.\cite{seabrook_high-throughput_2013, shi_dsf_2013} $T_{HE}$ is typicallly defined as the temperature where the DSF melting curve reaches a minimum of the first derivative, marked as $-\frac{dF}{dT}$. We will use this technique to determine $T_{HE}$ for HEWL in the presence and absence of added arginine.

Experimentally, three 25 $\mu$L replicate aliquots of each test sample were prepared as in Table ~\ref{tbl:si-samples} and pipetted into a 96-well PCR plate (Thermo Scientific). The experiment was then run using a CFS Connect RT-PCR instrument (Bio-Rad). Fluorescence intensity was collected over the range of 10\degree C to 95\degree C in 1\degree C increments. The total time for the experiment was approximately 103 minutes.

\pagebreak

\begin{figure}[H]
   \includegraphics[width=0.4\textwidth]{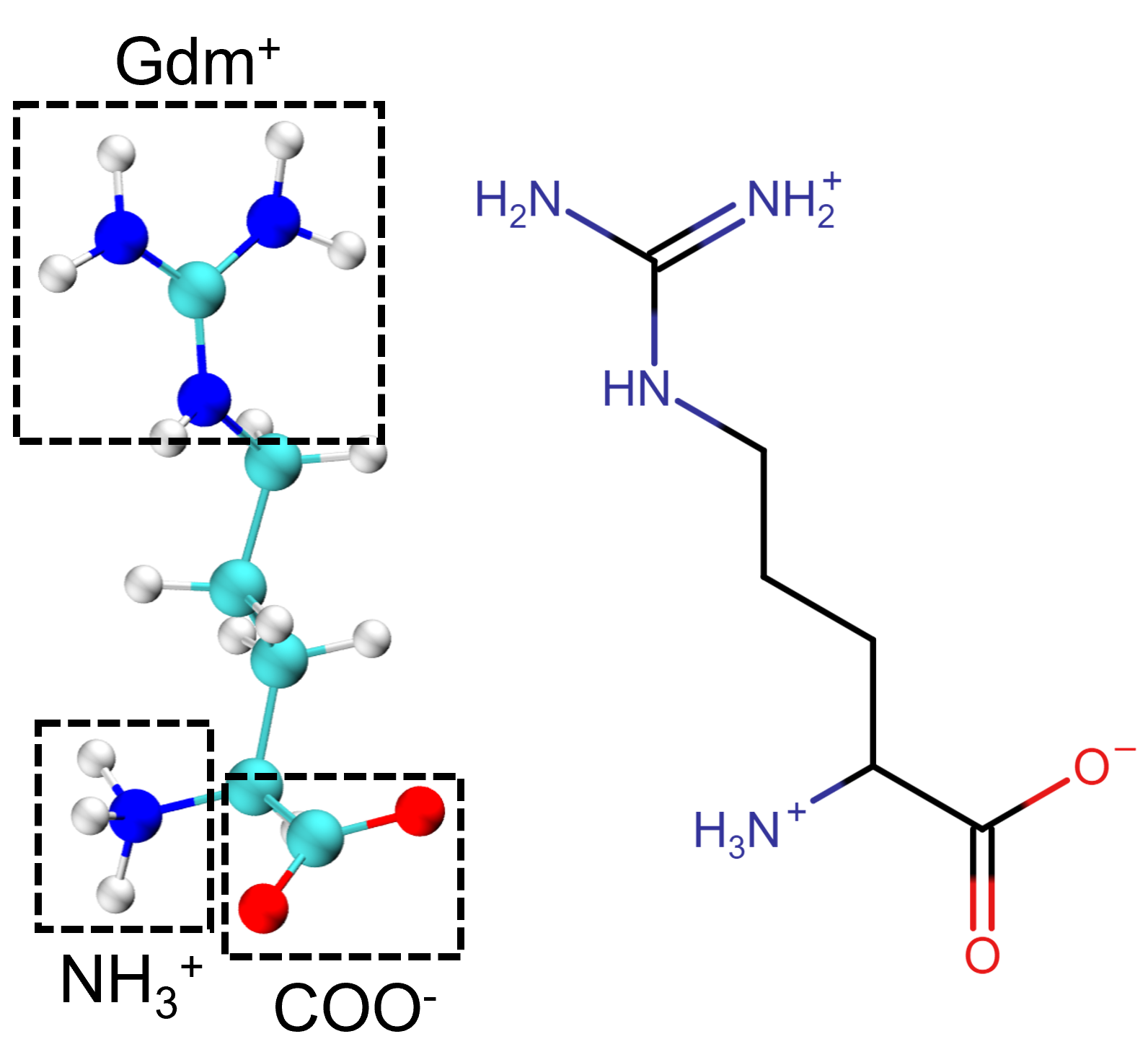}
    \caption{Representation of the structure of arginine. Boxes are drawn around the charged groups of arginine.}
    \label{fig:si-argstruct}
\end{figure}

\begin{figure*}[!ht]
   \includegraphics[width=0.9\textwidth]{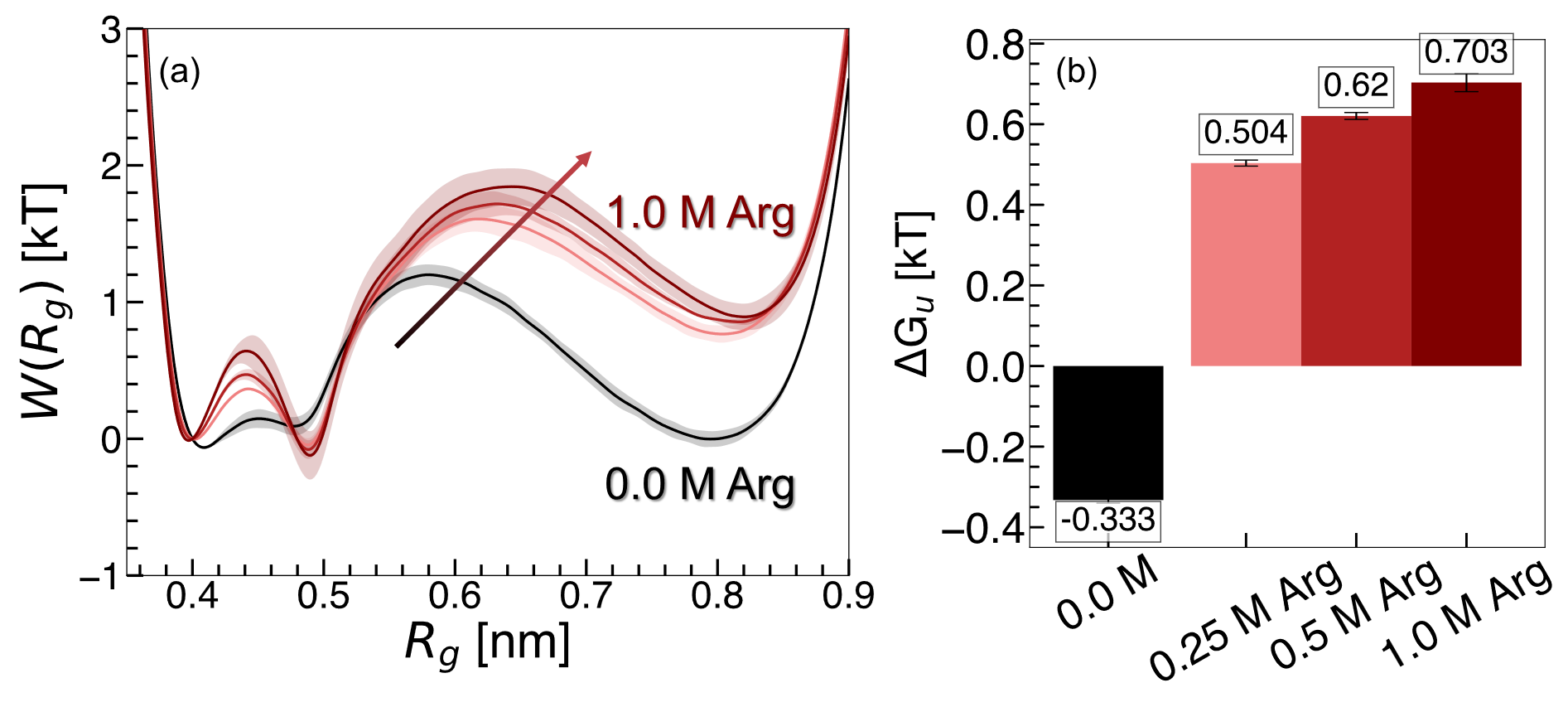}
    \caption{PMFs and free energies of unfolding obtained from REUS simulations. (a) PMFs of hydrophobic polymer along $R_g$ in pure water (black) and arginine (red) solutions. The arrow indicates the direction of increasing arginine concentration. All plots are normalized to 0 at $R_g = 0.4$ nm. (b) Free energies of hydrophobic polymer unfolding ($\Delta G_{u}$). Error bars were estimated as the standard deviation of PMFs obtained from three replicate simulations.}
    \label{fig:si-dbars}
\end{figure*}

\begin{figure*}[!ht]
\centering
   \includegraphics[width=0.9\textwidth]{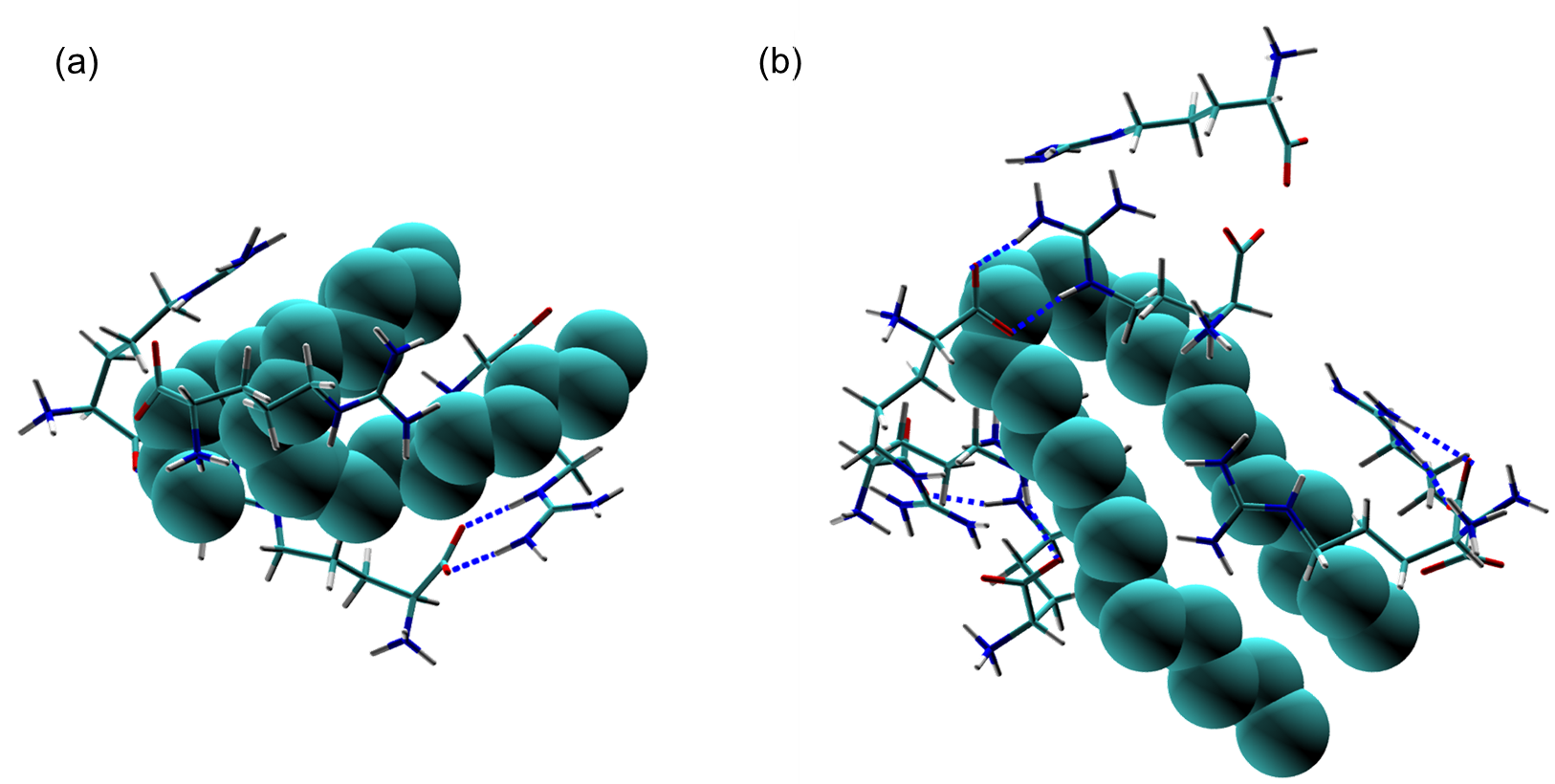}
   \caption{Representative snapshots of arginine encapsulating structures observed at the hydrophobic polymer surface. Snapshots extracted from the hydrophobic polymer in (a) unfolded and (b) folded REUS windows.} 
    \label{fig:si-argcage}
\end{figure*}

\begin{figure*}[!ht]
\centering
   \includegraphics[width=0.9\textwidth]{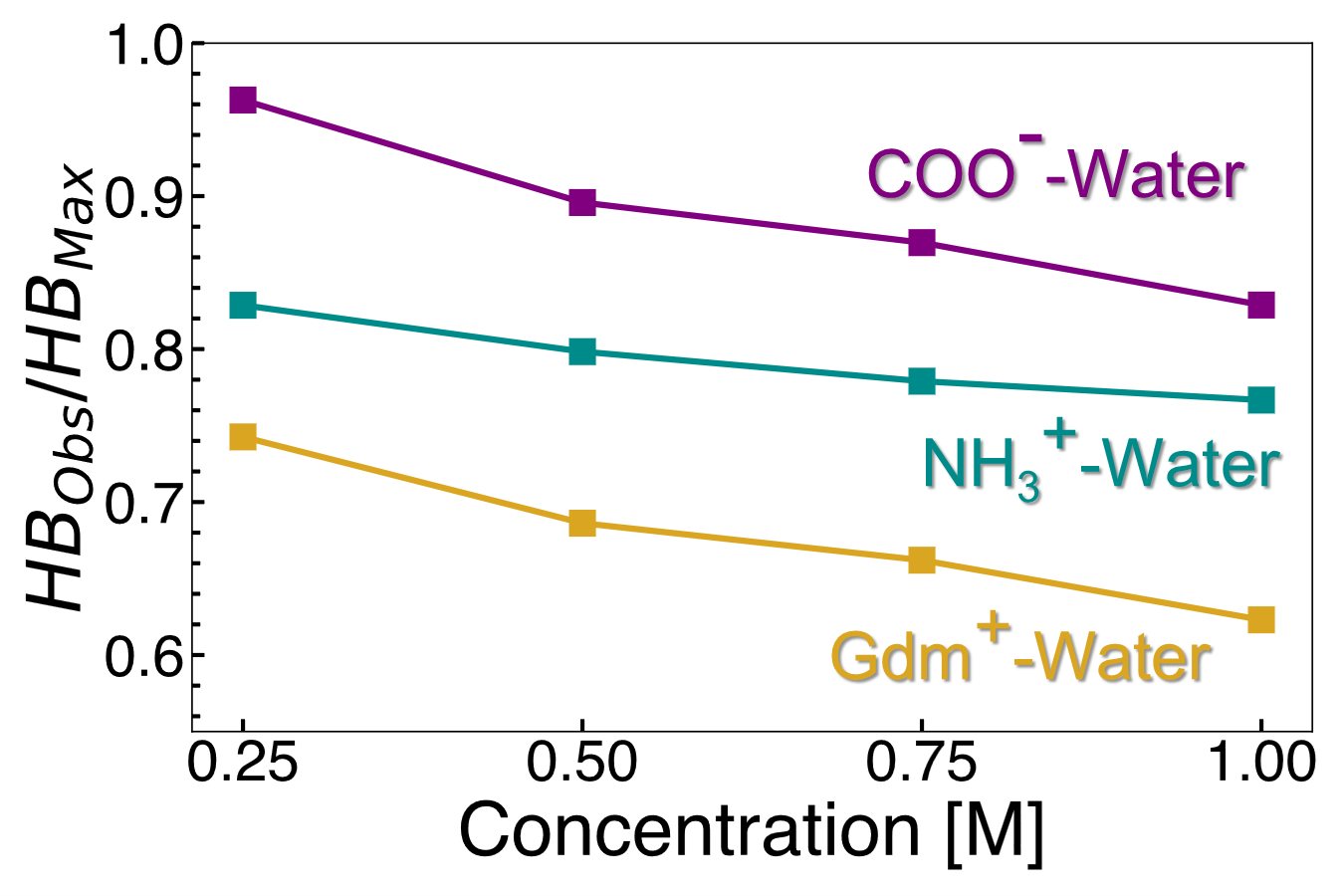}
   \caption{Fraction of observed hydrogen bonds ($HB_{Obs}$) relative to the maximum number of hydrogen bonds ($HB_{Max}$) per interaction group.} 
    \label{fig:si-hbond}
\end{figure*}

\begin{figure*}[!ht]
\centering
   \includegraphics[width=0.9\textwidth]{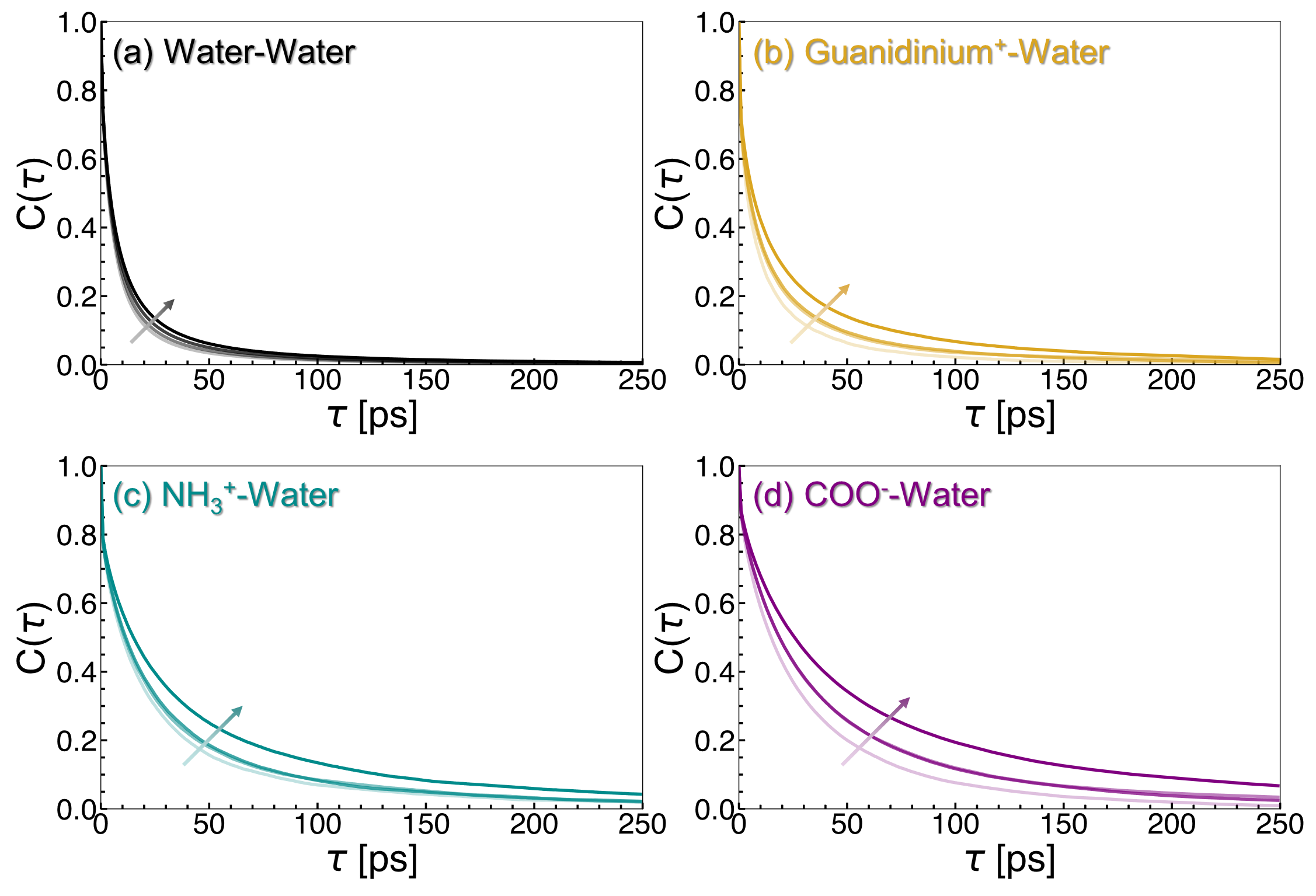}
   \caption{Hydrogen bond existence correlation functions for (a) water-water, (b) guanidinium$^{+}$-water, (c) NH$_{3}^{+}$-water, and (d) COO$^{-}$-water. Each plot is shown as a function of concentration, with increased shading (light to dark) denoting increasing arginine concentration.} 
    \label{fig:si-watCorr}
\end{figure*}

\begin{figure*}[!ht]
\centering
   \includegraphics[width=0.9\textwidth]{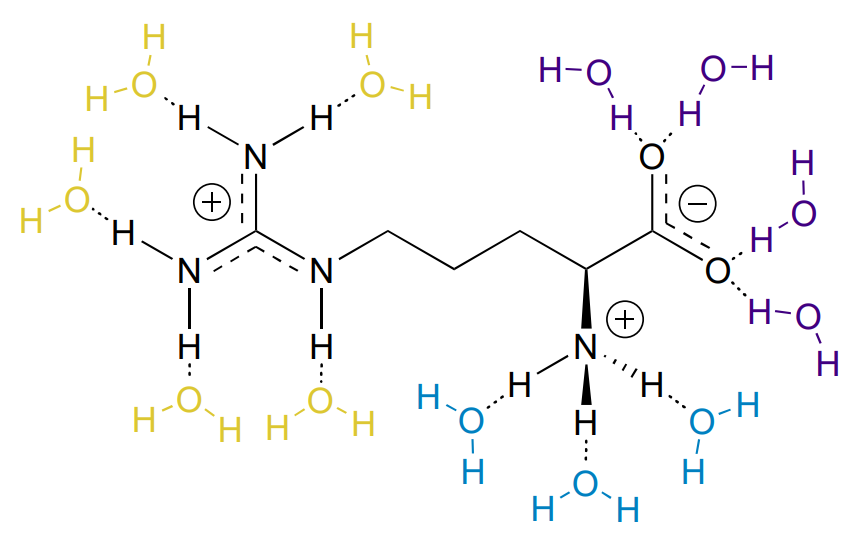}
   \caption{Illustration of arginine-water hydrogen bond interactions. Water molecules interacting with the Gdm$^{+}$ sidechain are highlighted in yellow, while those interacting with NH$_{3}^{+}$ and COO$^{-}$ are shaded in blue and purple, respectively.} 
    \label{fig:si-arg-hbonds}
\end{figure*}

\begin{figure*}[!ht]
\centering
   \includegraphics[width=0.9\textwidth]{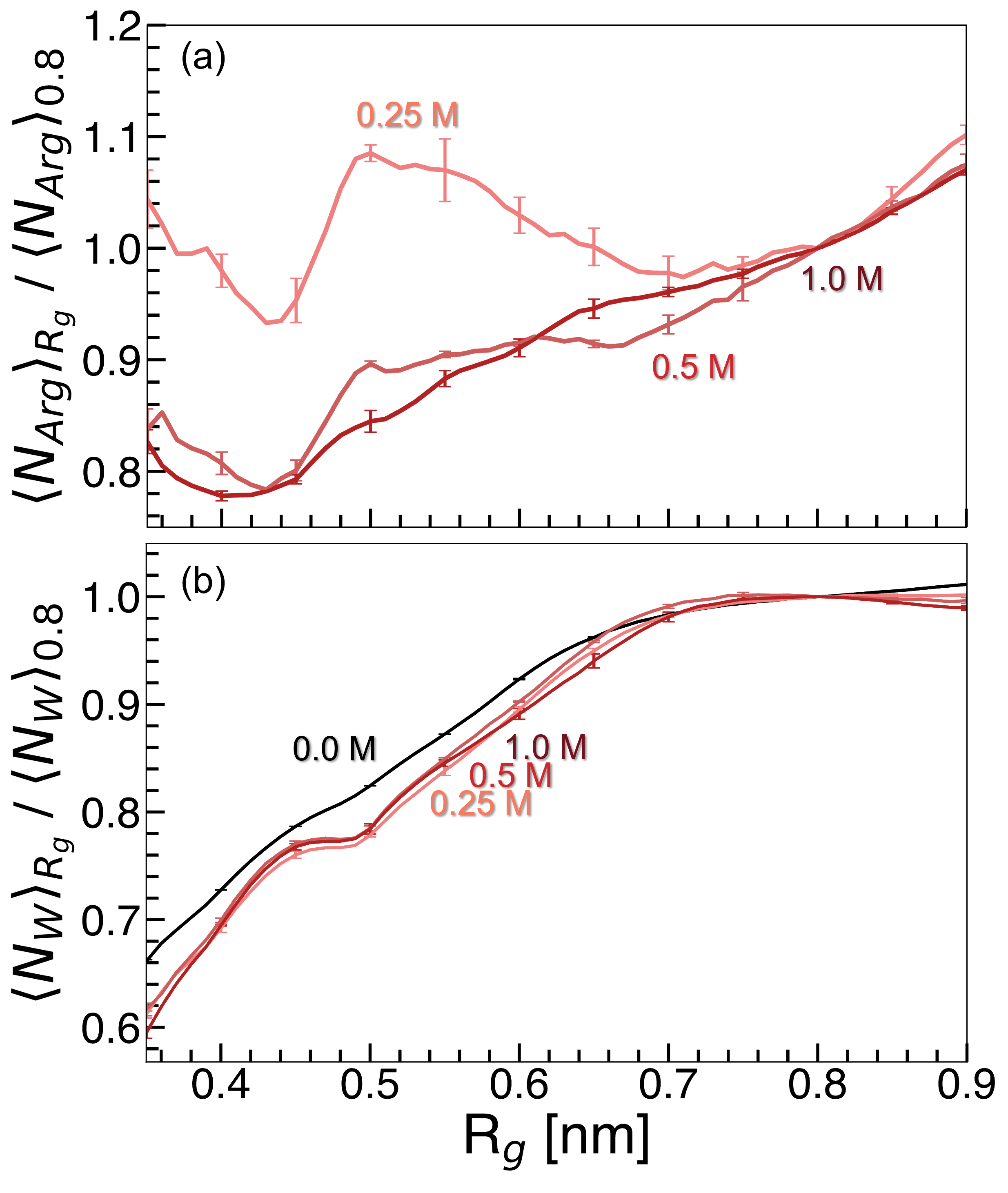}
   \caption{Quantification of arginine molecules and water molecules in the local domain of the hydrophobic polymer (within 0.5 nm). (a) Average number of arginine molecules in a given $R_{g}$ window. (b) Average number of water molecules in a given $R_{g}$ window. Values are normalized by the average value obtained at $R_{g} = 0$ nm. Means are estimated as the average value in a given bin for three replicate REUS simulations. Concentration is denoted by increased shading (light to dark).} 
    \label{fig:si-dewet}
\end{figure*}

\begin{figure*}[!ht]
\centering
   \includegraphics[width=0.9\textwidth]{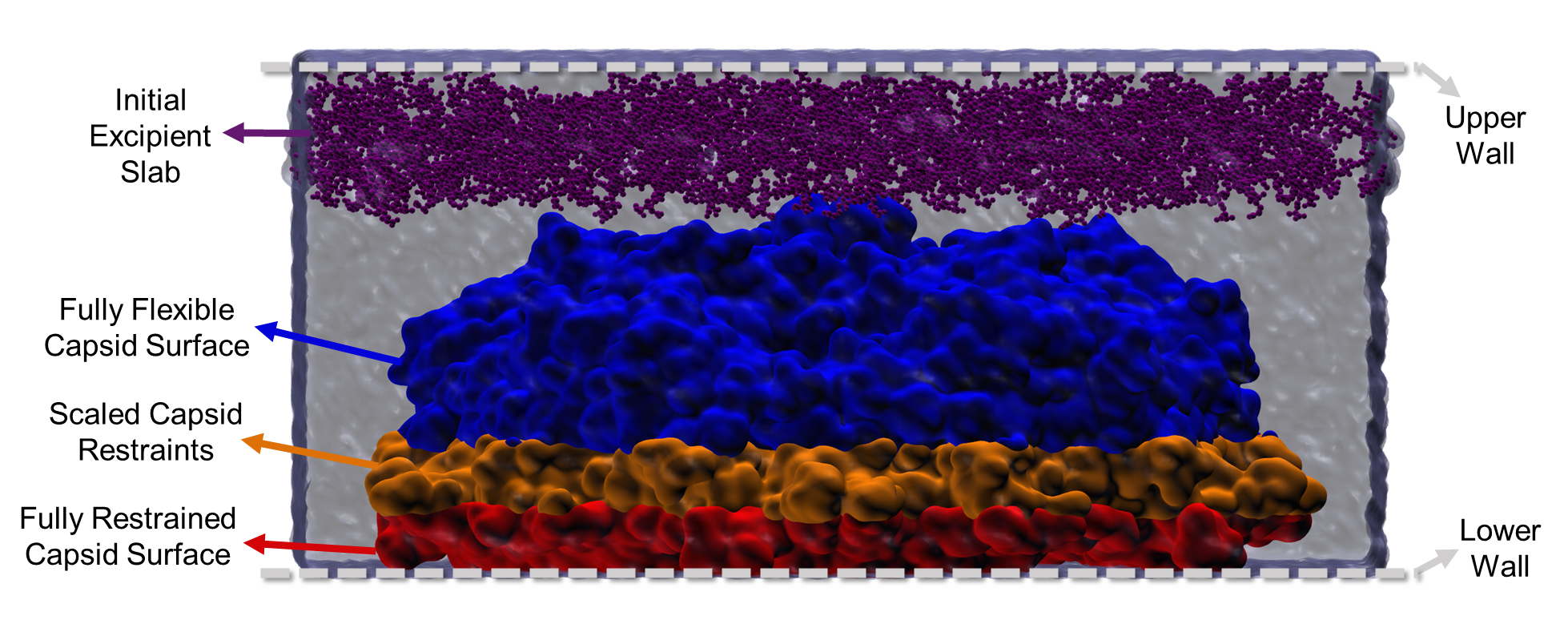}
   \caption{Schematic detailing the PPV 15-mer surface simulation.} 
    \label{fig:si-capsid}
\end{figure*}

\clearpage
\bibliography{ref}